\documentclass[aps,onecolumn,groupedaddress,floatfix,nofootinbib]{revtex4-1}

\usepackage[colorlinks=true,citecolor=red,filecolor=green,linkcolor=blue,pdfnewwindow=true]{hyperref}
\usepackage{amsmath} 
\usepackage{amssymb}
\usepackage[version=4]{mhchem}
\usepackage{color}
\usepackage{graphicx}
\usepackage{makeidx}
\usepackage{bm}
\usepackage[title]{appendix} 
\usepackage{geometry} 
\usepackage{float}
\usepackage{caption}
\usepackage{hyperref}
\usepackage[normalem]{ulem}
\usepackage{natbib}
\usepackage[dvipsnames]{xcolor}
\usepackage[english]{babel}
\usepackage[autostyle]{csquotes}
\usepackage{algorithm} 
\usepackage{algpseudocode}
\usepackage{epigraph}
\usepackage{multirow}

\hypersetup{urlcolor=blue}

\makeatletter
 
\newcommand{\Rmnum}[1]{\expandafter\@slowromancap\romannumeral #1@}
\makeatother

\begin{document}

\title{On study of transition fronts of Fisher-KPP type reaction-diffusion PDEs by non-linear transformations into exactly solvable class }

\author{Preet Mishra}
\author{Sapna Ratan Shah}
\author{R.~K.~Brojen Singh}
\email{brojen@jnu.ac.in}
\affiliation{School of Computational $\&$ Integrative Sciences, Jawaharlal Nehru University, New Delhi-110067, India.}

\begin{abstract}
{\noindent} Spatio-temporal dynamics of the evolution of population involving growth and diffusion processes can be modeled by class of partial diffusion equations (PDEs) known as reaction-diffusion systems. In this work, we developed a nonlinear transformations method that converts the original nonlinear Fisher-KPP class of PDEs into an exactly solvable class. We then demonstrated that the proposed nonlinear transformation method intrinsically preserves the  relaxation behavior of the solutions to asymptotic values of the non-linear dynamical system. We also show that these particular transforms are very amenable to yield an exact closed form solution in terms of the heat kernel and analytical approximations through the two variable Hermite polynomials. With this proposed method, we calculated the front velocity and shape of the propagating wave and showed how the non-linear transformation affects these parameters for both short and long epochs. As 
applications, we focus on solving pertinent cases of the Fisher-KPP type of PDEs relating to the evolutionary dynamics by assigning fitness to the mutant gene according to zygosity conditions. We calculated the relaxation of velocity with the parameters of the initial conditions in the following cases, namely, the Fisher, the heterozygote inferior fitness, the heterozygote superior fitness, and finally a general nonlinearity case. We also verified previous conjectures through the exact solutions computed using the proposed method.   \\

{\noindent}{{\it \textbf{Keywords:}} \textbf{Non-linear dynamics}; \textbf{Fisher-KPP type PDE}; \textbf{Transformation methods}; \textbf{Lower and Upper Bounds}; \textbf{Population genetics}. }

\end{abstract}

\maketitle

\tableofcontents

\section{Preliminaries and Known Results}
{\noindent}One of the fundamental origins of the spatiotemporal patterns in genetic diversity in evolutionary process in a certain landscape is the complex dynamics due to random interaction among the species which follow basic genetic laws, and is driven by variations surrounding them \cite{Feder,Sork}. There have been attempts to understand these dynamics, but still it is far from resolving many fundamental issues, for example, mechanisms in gene flow \cite{Lenormand}, crucial aspects of fitness \cite{Orr}, dynamics of adaptability and selection in mutation phenomena \cite{Sniegowski,Barrett}, accurate mapping of phenotypic patterns by genotype dynamics \cite{Barrick}, and many others. These issues are multidimensional and thus the need to address them from mathematical models. We review some of the important works as discussed below in order to address our work.

\subsection{Role of diffusion and growth in genetics}
{\noindent}Systems approach to study the processes involving evolution of new mutants, their growth by multiplicative processes and their dispersal in a spatial habitats results in complex nonlinear dynamical systems  with very many degrees of freedom to describe them. Broadly, modeling the temporal dynamics (both discrete and continuous) of these systems together with the spatial dimension occupies a pivotal role in describing qualitative and quantitative understanding of the existence of genetic diversity \cite{kimura_book,moran}. This approach, which addresses these issues/problems, describes what is commonly known as gene flow dynamics or clines \cite{b1,b2,b3}. Further, migration or dispersal is always a causative agent of the dynamics in the population \cite{n1,n2}. This factor also effects the way genes are distributed in a habitat along the evolutionary process \cite{b3,arms}. Now, several of these models have also been employed to generate predictive and tackle real-world control problems of complex genetic diversity \cite{b4}. \\

{\noindent}Prior theoretical works on spatial distribution \cite{kpp,fisher,kendall,skellam} of species evolution and the interplay of genetic differentiation has not only been given mathematical and statistical rigour but also advanced techniques of understanding dynamical systems (ordinary differential equations (ODE)s, partial differential equations (PDE)s) \cite{murray}. In general, these mean-field continuum models described by ODEs or PDEs are formulated as scaling limits of some discrete models. A beautiful illustration of the microscopic origin of such mean-field equations are given in \cite{derrida} and a detailed review of this given in an excellent review in \cite{panja}. In these models, the amount (density) of entity is generally represented by a smooth function $u : \mathbb{R}^d \times [0,\infty) \rightarrow [0,1]$. The dynamics of growth and dispersal of the entity is then given by a nonlinear partial differential equation governing the rules of change in the density over spatial and temporal dimensions in the following form,
\begin{eqnarray}
\label{eq1}
\frac{\partial{u}}{\partial{t}}= D\nabla^2 u + F(u,x,t)
\label{rde}
\end{eqnarray}
where,  $\nabla^2$ in equation \eqref{eq1} is the Laplacian operator for dimension $d=1 $ and $F(u,x,t)$ denotes the function for all interactive mechanisms of the system under consideration. It is $F$ that contains the non-linearity contributions and therefore provides various complex dynamics, states and patterns evolved with the system.\\

{\noindent}Though there are obvious merits and drawbacks of such field theoretic treatment of the real phenomenon, these equations arise in many phenomenological problems such as transport equations, physico(mechano)-chemical models of morphogenesis, spatially oscillating chemical reactions (BZ reactions), conduction of nerve impulses, combustion theory etc which can be found in the excellent examples and references in \cite{murray}. We would like to state that there is a huge literature on the methods and the universality features for these classes of PDEs, and hence, we have taken a very small subset of the methods used to study our problem. In this work no numerical integration of the PDE were done and we have relied only on analytical methods and basic graph fitting algorithms to derive all the results.

\subsection{The Fisher-KPP class of PDE}

{\noindent}Consider the following initial value problem of the equation \eqref{eq1} as stated as in the following,
\begin{eqnarray}
\label{eq2}
u_{t}(\Bar{x},\Bar{t})&=&D u_{\Bar{x} \Bar{x}}(\Bar{x},\Bar{t})+kF[u(\Bar{x},\Bar{t})]\\
u(\Bar{x},0) &=& \phi(\Bar{x})\nonumber
\end{eqnarray}
By defining a suitable scaling transformation \cite{murray} of the variables $(\Bar{x},\Bar{t})\rightarrow (x,t)$  as given below,
\begin{eqnarray}
\label{eq3}
t=k\Bar{t}\quad \text{and} \quad x=\Bar{x}\sqrt{\frac{k}{D}}
\end{eqnarray}
The equation \eqref{eq2} can be written in terms of the new variables as in the following,
\begin{equation}
\label{fkpp}
u_{t}(x,t)=u_{xx}(x,t)+F(u(x,t))
\end{equation}
Fisher took a specific form of $F$ as $F=u(1-u)$ and studied the above equation \eqref{fkpp} extensively as a travelling wave solutions and their properties, which in now known as Fisher waves \cite{fisher} also KPP et al \cite{kpp} used the form $F= u(1-u)^2$ . The central goal of this paper is to calculate explicit closed form bounds of the equation \eqref{fkpp} which show a similar behaviour as close to the original one.

\subsection{Mathematical techniques to study Fisher-KPP type equations}

{\noindent}The Fisher-KPP type PDEs for general form of $F(u)$ are intractable except for specific simple forms of $F(u)$. In this work, we restrict ourselves to the classes of Fisher-KPP type \cite{fisher,kpp} of reaction diffusion PDEs as represented by \eqref{fkpp} and provide a brief overview of the analytical results obtained in this class of PDEs. There have been various classic prior works \cite{aronson1,aronson2} which have resulted in the widely known exact solutions to some of the functional forms of the $F(u)$ in \eqref{fkpp} . To cite a few, one can see the excellent results for exactly solvable class given in \cite{kametaka, zepetella, bramson1, wang, fuchs,brunet2015}. All of the above cited works give general idea about the solution spaces of such PDE for a specific class of non-linearities in $F(u)$. An all time classic review of these results can be found in \cite{saarloos1}.\\

{\noindent}During the last few decades, some of the major fruitful and powerful methods for studying the solutions of Fisher-KPP class of PDEs have been reported, where, these works are to rigorous asymptotic analysis \cite{murray1, gcole, larson, rothe, sheratt, needham1, needham2}, norm-estimates as done in \cite{hamel1,hamel2,hamel3,moet,sattinger}, perturbation-expansion and Renormalization Group (RG) approaches \cite{rosales,puri1,dixon,puri2,bricmont}, and more recent works on representing these PDE solutions as global attractor problems using renormalisation group theory approach can be found in \cite{ tang2,fischer,tang1}. \\ 
 
{\noindent}Further, the role of initial conditions on the effects on the Fisher wave front propagation of Fisher-KPP type PDEs has been studied extensively with interesting results. Some of the excellent results worth mention are, first, a beautiful result in (Theorem 6.3 \cite{sattinger}) which essentially asks about stability conditions, how far is the initial data from a travelling wave profile and then gathers the information about the solutions depicting KPP wave stability conditions. There are also results that the weighted norms of the solutions of the \eqref{fkpp} are bounded (Theorem 1 \cite{moet}) based on $L_p$ norms of the initial data. All these interesting results have motivated us to investigate deeper insight to the hypothesis of the existence of the properties of the nonlinear transform that has been thoroughly studied in this work. In the next section, we give some more rigorous rationale behind the work.

\subsection{On non-linear transformation techniques }

{\noindent}The transformation techniques to deal with Fisher-KPP type PDEs are deeply in the spirit of the Lie invariances. In this approach, we try to find nonlinear transformations by matching terms that leave the original PDE invariant or very close to it. One of the example is the Cole Program to generate similarity solution of the heat equation \cite{weizsacker,cole1,cole2}. The Renormalisation Group (RG) method for the PDEs \cite{bricmont} builds up on the scaling properties to deal with the model PDE and study critical phenomena of the model system. Critical exponents extracted from the fixed points of this iterative technique gives the solution spaces of the model system. Below we give a brief outline as to how we can synthesize these methods together with non-linear transformations to yield closed form solutions to a class of PDEs. \\

{\noindent}The methods mentioned above, which can be applied to Fisher-KPP type equation, and on the approximations using first order terms in the perturbation expansion \cite{rosales} give us the possibility to study interesting properties and deeper insights of its solution spaces. We will later see that the results of the usage of these techniques \cite{puri1,puri2} are found to be as a special case of the transformation function which we will be discussing below. Extensive work on this can be found in \cite{bricmont}. We build our work on these observations of what happens to self similar solutions if they are under some nonlinear transformations with some well defined properties like monotonicity etc. This put our work predominantly in the class of Lyapunov type of methods. In this direction, there have been few previous works regarding Lyapunov functions in the solution manifolds of PDEs \cite{powell} on the style of transformations which involves a minimum of one arbitrary function, for example, using only the classical heat polynomials. However, increase in the number of such functions does not guarantee any closed form solutions as is evident in the Painleve analysis (section III in \cite{powell}) but it does show the behaviour due to some simple nonlinearities in the function. Hence we, have been motivated in this work to first check the limits of this kind of transformation equations involving minimal number of approximating functions. \\

{\noindent}Further, we not only want to calculate specific forms of the solutions but also to preserve the variety of structures present in the original nonlinear problems like the relaxation of the front velocities etc. Complementary to the above mentioned methods, another powerful method is that of the transformation of solutions via nonlinear ones into kinds of equations that can be exactly solvable in closed form. This kind of program is the main goal of this paper. The work in \cite{thomas,majumdar} can be referred to as a beginning of these transformation techniques to deal with the nonlinear PDEs. This program was heavily used by Montroll  (in Appendix I of \cite{montrollcascade}) who cites many original papers therein regarding the use of such transformation techniques. Some interesting transformations for the nonlinear PDEs are also discussed in \cite{montroll,west, weiss,dixon,reid}. In the present work, we follow the technical ingredients of the program initiated in the pioneer papers of Montroll \cite{montrollcascade,montroll}, where, he heuristically established how the growth and diffusion PDE problems can be transformed into the problem of studying solutions of known PDEs. We thus extend and complete the analysis of the behaviour of closed form solutions obtained through these kind of transformations.\\

{\noindent}Hence, this work is more in the lines of a phenomenological approach where a priori knowledge of the entity under study dictates the kind of solution spaces to be used and their subsequent behaviour. We would also like to point out that the performance of this method relies on how well the transformation preserves the initial condition and boundary conditions as well. The motivation of the paper is then the observation that a wide class of simpler reaction-diffusion PDEs can be tackled through such transformations efficiently. This serves as a two way goal such as of providing deep insights and also faster computing of the original problem.\\ 

{\noindent}The structure of the paper is as follows. We begin by stating a general transformation function that maps the nonlinear PDE solution into a solvable class. Then, we introduce two type of ansatz, first one involving homogeneous heat equations whose solutions are known extensively, and, second one involving diffusion PDEs whose solutions are also known in terms of the heat equations.  We then proceed to show some examples from population genetics as an application for the method developed. We then draw conclusions based on the results we obtained. 

\section{Bounds of a class of Fisher-KPP type equations}

{\noindent}The well established results about this PDE \eqref{fkpp} with $F$ having special properties is that the solutions represent saturation conditions over the entire spatial dimensions as time moves forward. Montroll \cite{montrollcascade} was the first to recognise this phenomenon and devised a nonlinear transformation that yields good approximations to the original PDE. We follow his technical procedure, extend his results and also obtain some new transformations of the same class giving new insights in the results. \\

{\noindent}We would like to point out that this program will not aim to provide explicit solution to the original PDE but provide solutions that remain asymptotically bounded above or below the original solution and become equal to the original solution in limiting case. We have two types of ansatz :\begin{itemize}
    \item Transformation explicitly depending on the nonlinearity $F(u)$ and the homogenous heat equation this we call type -I
    \item Transformation that involve non-homogeneous but solvable PDE we call type -II ansatz
\end{itemize}

{\noindent}We begin by defining a class of non-linear transformation:

\noindent\textbf{Definition :} \textit{G  maps $(0,1) \to \mathbb{R}^+$ . It has the following properties: it must be monotone (increasing or decreasing) and there must exist $G^{-1}$ that maps the $\mathbb{R}^+ \to (0,1)$. }

\noindent Using the above definition and two subsidiary functions whose properties we can control, we can construct a nonlinear transformation that can yield solutions to our original PDE. We use the following G :
$G\big[u(x,t)\big] = f(t)h(x,t)$ where $u(x,t)$ is the solution of the reaction diffusion PDE \eqref{fkpp}. the two functions $f(t)$ and $h(x,t)$ are derived according to the problem. 

{\noindent}The function defined above is quite similar to a Lyapunov type of function generally involved in the study of dynamical systems and we will show it in the succeeding section. The problem we considered in this section is to study the behaviour of a nonlinear transformation function. The analytical behaviour of the transformation function comes from the multiplicative type of growth and diffusion of an entity that can grow using a myriad of processes like autocatalysis, self-renewing etc in various dynamical systems \cite{frank}. Further, the observation that a saturation type in the growth of the function leads to the following classes of nonlinear PDEs was harnessed from the beginning analysis by Fisher and KPP et.al.\cite{fisher,kpp}.\\

\begin{figure}
    \centering
    \includegraphics[width=0.6\linewidth]{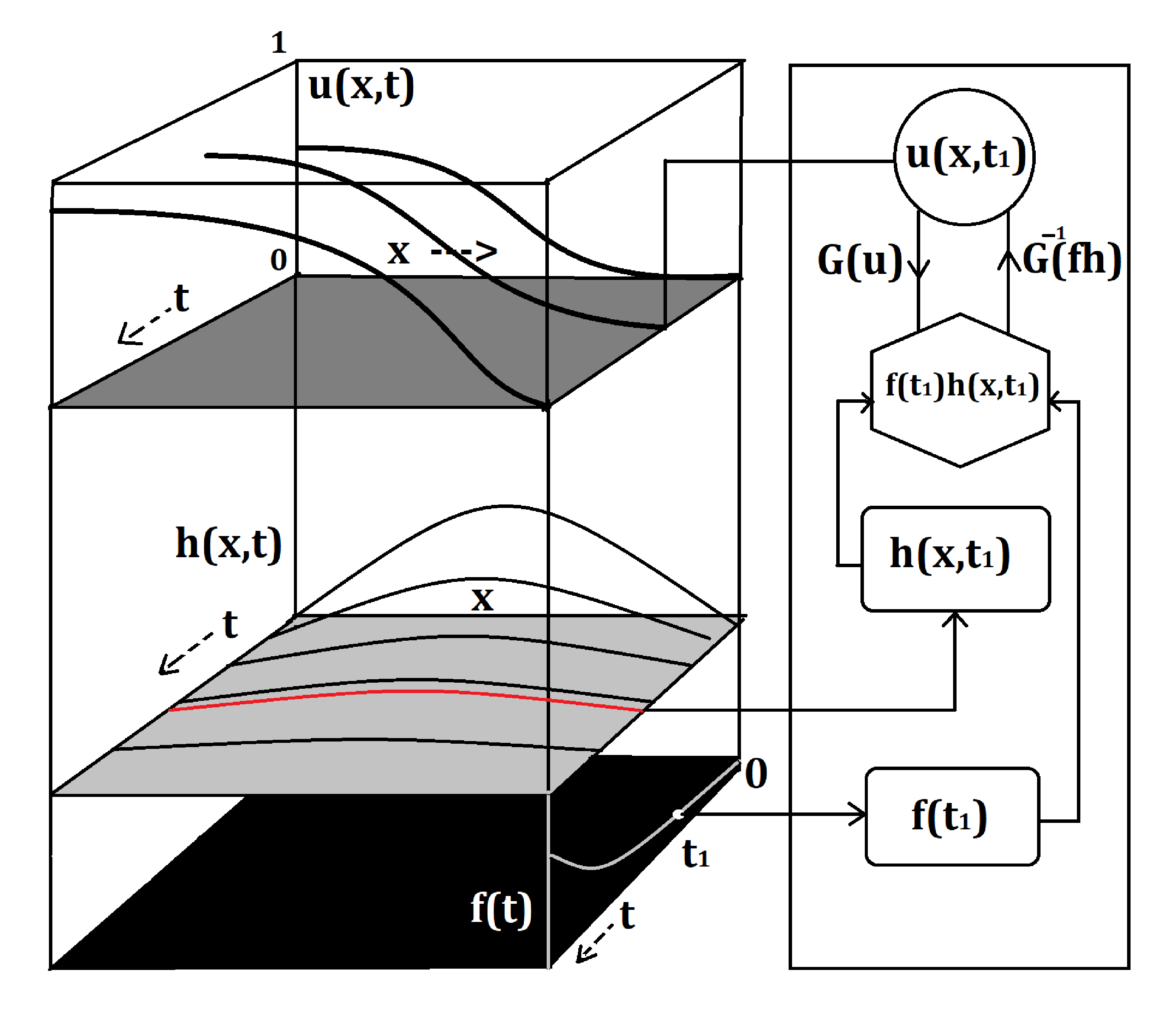}
    \caption{A representative schema for the transformation function.  }
    \label{fig1}
\end{figure}
{\noindent}To understand this technical program in a heuristic manner, we can think of the following properties of the variables and functional forms involved as given in Fig. \ref{fig1}. We know from the phenomenon modeled by these PDEs and also from the extensive prior works that the exact qualitative nature of how the solutions of $u(x,t)$ depends on the initial data behaviour $\forall x \in \mathbb{R}$ and the behaviour of $F(u) \forall u \in [0,1]$. We also know that there exists a value $\theta$ which is a stable fixed point of the map $F(u)$. Thus, we can understand that the solution $u(x,t)$ of the PDE \eqref{fkpp} generally converges to the value $\theta \in [0,1]$ in the form of a travelling wave $\psi(x+ct)$ (Theorem 13 of \cite{kpp}). In these studies, what we do not know so far is an explicit closed form of the function $\psi(x+ct)$. Hence, our goal is to construct tight approximate closed form bounds for this original function $\psi$ .\\

\noindent Our central hypothesis for this $G^{-1}$ is to look for the following property,
\begin{eqnarray} 
\lim\limits_{t \to \infty} G^{-1}\left[f(t)h(x,t)\right] \to \theta
\end{eqnarray}
uniformly for all $x$ in the the form of a travelling wave $\psi$. Since we know the exact closed forms of $h(x,t) $ and $f(t)$, we can then write the solution $\psi$ in a closed form from $G^{-1}$.

\subsection{Transformation ansatz type-I}


{\noindent}Let us consider the following transformation functions $G$ which combines both growth and diffusion as given by,
\begin{eqnarray}
G\big[u(x,t)\big] = f(t)h(x,t)
\label{trans}
\end{eqnarray}
where, $h(x,t)$ satisfies a homogeneous heat equation given by,
\begin{equation}
h_t=h_{xx}
\label{heatf2}
\end{equation} 
and $G$ has to be determined from the non-linear function $F(u)$. The initial condition of this equation \eqref{heatf2} with the transformation equation \eqref{trans} is given by, $h(x,0)=G\left(u(x,0)\right)=G\left[g(x)\right] $, where, $u(x, 0)=g(x) \in [0,1]  \; \forall \;\; x \in \mathbb{R}$ and $f(0)=1$. Now, from the above transformation \eqref{trans} and the heat equation \eqref{heatf2}, we arrived at the following equation,
\begin{eqnarray}
\label{traneq}
u_{t}(x,t)=\bigg[  u_{x x}(x,t)+u^{2}_{x}(x,t)\frac{G^{\prime \prime}(u)}{G^{\prime}(u)}  \bigg ]+\frac{G(u)f'(t)}{G^{\prime}(u)f(t)} 
\end{eqnarray}
where $G'(u) =\frac{dG(u)}{du}$ and $G''(u) =\frac{d^2G(u)}{du^2}$ and so on. Further, from the relation between $F$ and $G$ given above, the PDE \eqref{traneq} becomes,
\begin{eqnarray}
u_{t}(x,t)=u_{x x}(x,t)+(u_{x}(x,t))^{2}\left[\frac{\frac{f'(t)}{f(t)}-F^{\prime}(u(x,t))}{F(u(x,t))}\right]+ F(u(x,t))
\label{meq}
\end{eqnarray}

The rationale behind this type of transformation is that we can get two scenarios in the results, both of which are fruitful in explaining the dynamics of the system. First, if the transformation leaves the original PDE invariant, we have solved our problem exactly as we know all the properties of the functions ($h(x,t) \text{and} f(t)$) used to construct the original variables ($u(x,t)$). Second, if the transformation introduces extra terms, we get a surrogate PDE which we can modulate by the choice we have in the two free functions ($h(x,t) \text{and} f(t)$) . The schematic diagram of the proposed technique of such ansatz is given in the Figure \ref{fig1}.\\

{\noindent}Now, comparing the equations \eqref{fkpp} and \eqref{traneq}, and taking the terms that do not involve derivatives of $u$ in both the equations, we get:
\begin{eqnarray}
G\left[u(x,t)\right]=\exp \left[\frac{f'(t)}{f(t)}\int \frac{du}{F(u)}  \right] 
\label{geqn}
\end{eqnarray}
From this comparison, if we take $\frac{f'}{f} = 1$, we get an important quantity given by,
\begin{eqnarray}
\frac{d}{du}\big[\ln G(u)\big]  = \frac{1}{F(u)}
\end{eqnarray}
This result indicates how the magnitude of the overall rate of change of the function $G(u)$ with respect to the variable $u$ i.e. $\frac{G^\prime}{G}$ inversely related to the non-linear function $F(u)$. Moreover, the monotonicity nature of $G(u)$ such that this function's inverse $G^{-1}(u)$ may be taken as a kind of flow that leads to a steady state solution as we evolve the initial condition with some kernel in our case of the heat kernel equation. 

\noindent Thus we have reduced the studying the above nonlinear PDE \eqref{fkpp} with nonlinearity $F(u)$ into that of studying the following Integral equation:
\begin{eqnarray}
\exp \left[\frac{f'(t)}{f(t)}\int \frac{du}{F(u)}  \right]=\left[\frac{f(t)}{f(0)\sqrt{4 \pi t}} \int_{\mathbb{R}} G(u(x',0)) e^{- \frac{(x-x')^2}{4t} } dx'\right] 
\end{eqnarray}
where the $f(t)$, $u(x,0)$ are all known a-priori and G is known from \eqref{geqn}. The properties that make heat kernel an excellent choice is the norm estimates that are extensively known. We cite one of these a-priori estimate in the Appendix A (subsection 1). We have also provided in Appendix A(subsection 2) the methods of evaluating the convolution of the initial conditions with the heat kernel and in Appendix A (subsection 3) a power series solution representation using the two variable Hermite polynomials.

\noindent\textbf{Proposition 1 :} \textit{ The solutions of 
\begin{equation}
\label{trfeq}
    v_t(x,t) = v_{xx}(x,t) + \left[ \frac{\frac{f'(t)}{f(t)} - F'(v(x,t)) }{F(v(x,t))} \right] (v_x(x,t))^2 + F(v(x,t)) 
\end{equation}  yields tight closed form bounded approximations of
\begin{equation}
\label{fk}
    u_{t}(x,t) = u_{xx}(x,t)+F(u(x,t)) 
\end{equation} $\forall (x,t) \in \mathbb{R} \times [0,T]$  given that \quad  $f(t) \neq 0  \quad \text{at} \quad t=0$ and $\left( \frac{\frac{f'(t)}{f(t)} - F'(v(x,t)) }{F(v(x,t))} \right) (v_x(x,t))^2$  remain uniformly bounded in $\mathbb{R} \times [0,T]$ set. }\\

\noindent \textbf{Proof:} The detailed proof of the proposition can be found using the maximum principle and parabolic inequalities (as given in Proposition 2.1 of \cite{aronson1} p.12) and it is proved that the solution class is a upper bound for a specific case of $F(u) = u(1-u)$ is given in \cite{larson} .\\

\subsubsection{Dispersion relations and Role of Growth Functions f(t) }

{\noindent}Consider the PDE \eqref{meq} for the linearised case, $F(u) = u$, then at the front of the wave, we have the dominant contribution to the growth as given by, 
\begin{equation}
\label{lieq}
u_t = u_{xx} + \left[ \frac{\frac{f'(t)}{f(t)} - 1 }{u} \right] u_x^2 + u 
\end{equation}
Following the asymptotic analysis method given in \cite{murray}, we perform the leading edge analysis. We assume an initial condition of the following form, 
\begin{eqnarray}
u(x,0) \sim e^{-ax} \quad \text{as} \quad x\rightarrow \infty 
\end{eqnarray}
Now, on taking the travelling wave ansatz with $u(x,t) =u(x-ct)$ of the form $A e^{-a(x - ct)} $, and then substituting in equation \eqref{lieq}, we can obtain the relation between $c$ and $a$ as,
\begin{eqnarray}
c =  a \left[ \frac{f'(t)}{f(t)} \right] + \frac{1}{a}
\end{eqnarray}
Now, if we take exponentially growth function, $f(t)= e^t$, we get $\frac{f'(t)}{f(t)} = 1$, we get the dispersion relation as $c =  a  + \frac{1}{a}$. This result verifies our ansatz that the front of the solution of our transformed PDE follows that of a linearized one.

\subsubsection{Effects of transformation function on the relaxation of transition fronts}

{\noindent}The front solution has been itself a topic of various studies. Two of the major attributes one studies is the shape and the velocity of the front. In the following section, we demonstrate the properties of the transformation function inherited from the the non-linearity of the function $F(u)$.\\

\noindent\textbf{Fisher Case : $F(u)=u(1-u)$}\\

\noindent This function was used by Fisher \cite{fisher} and was the only one treated by Montroll \cite{montroll} who obtained certain analytical estimates of the velocity of the saturation front only. We complement and extend his results by analysing the behaviour of solutions and obtaining the velocity estimates. Now, from the function, $F(u)= u(1-u)$, we have, $F^\prime(u) = 1-2u$. Then, substituting these $F(u)$ and $F^\prime(u)$ to the PDE \eqref{meq} we get:
\begin{eqnarray}
\label{c1}
u_{t}=u_{x x}+u_{x}^{2}\left[\frac{\frac{f'(t)}{f(t)}-1+2u}{u(1-u)}\right] + u(1-u)
\end{eqnarray} 
the transformation equation \eqref{geqn} with exponential growth function, $f(t) =  e^t$ becomes,

\begin{eqnarray}
G(u) = \frac{u}{1-u} = e^t h(x,t) 
\end{eqnarray}
such that one can get by the inverse transformation $u$ as:
\begin{eqnarray} 
u = \frac{e^t h(x,t)}{1+e^t h(x,t)}
\end{eqnarray}

\noindent We would like to point out that this solution was also was recorded by \cite{puri1} and ( Eq.4 in \cite{puri2} p.454) through completely different methods. This suggests that there are deep connections between these transformations and the perturbation expansion methods. \\

{\noindent}We further calculate the following quantities following Fisher in (Section 8 of \cite{fisher}) as :
\begin{itemize}
    \item Mode of heterozygote distribution, which represents the point of equality of the gene ratios.
    \item Median of the heterozygote distribution.
\end{itemize}

\begin{figure}
    \centering
    \includegraphics[width=1.0\linewidth]{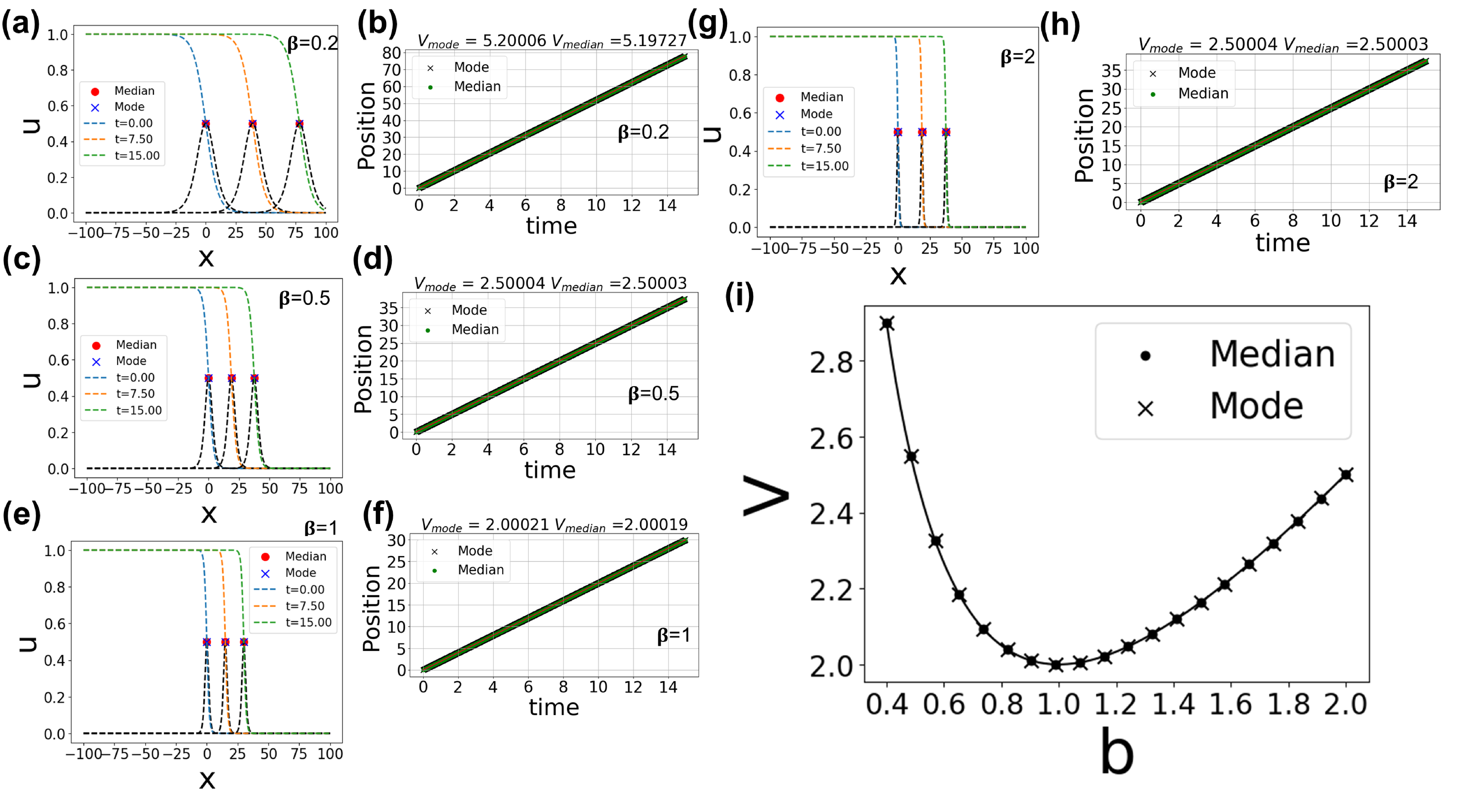}
    \caption{Graphical representations for the exact solutions pertaining to the to the sigmoidal type of initial condition shape controlled by parameter $b$ with $F(u) = u(1-u)$. Waves progression and slopes giving the front velocity  by tracking the mean and median of the heterozygote distribution for  (a)-(b)$b=0.2$ (c)-(d) $b=0.5$ (e)-(f)$b=1$ (g)-(h)$b=2$ . (f) Dispersion relation for the wave speed ($V$) with the initial condition shape control parameter $b$  }
    \label{fig2}
\end{figure}

{\noindent}We follow Fisher's method of calculating the velocity of the front and compare it with the asymptotic analysis of the solutions obtained. We report the following new results. We take first an initial condition that represents the condition that the mutant population exists in the entire negative half line, and they have advanced in $b$ units ahead of the origin as shown in Figure \ref{fig2}(a-h). Thus, this gives us the initial function in the form $u(x,0)=\frac{1}{1+e^{b x}}$. This implies that $u(x,0) \sim e^{-b x}$ as $x \to \infty$, and hence, the velocity of the wave should be governed by the behaviour of $u(x,0)$ at $x\to\infty$ which is given as  $ V = b+\frac{1}{b} $. We further take $f(t) = e^t$ and solve the transformation equation \eqref{meq}. The transformation gives the exact solution as,
\begin{eqnarray}
h(x,t) &=& e^{-b(x-b t)}\\ 
\implies u(x,t)&=&\frac{1}{1+e^{b(x-(b+\frac{1}{b})t) }}
\label{wf}
\end{eqnarray}
Further, the rate at which the mutant front propagates is given by when the term in the exponential vanishes. This condition is given by, $x=(b+\frac{1}{b})t$, from which one can obtain the rate at which mutant wave propagates as given by,
\begin{eqnarray}
\frac{dx}{dt}=b+\frac{1}{b}
\label{vf}
\end{eqnarray} 
We again verify this velocity by using the mode and the median of the heterozygote distribution as given in Figure \ref{fig2}(i). Thus we verify this results by using the method given by Fisher which also shows very good accuracy of calculating the front solutions by discretely evaluating the function $u(x,t)$ with reasonable spatial and temporal resolutions ($\delta x=0.04, \delta t = 0.0015 $). \\

{\noindent}Now, we take another initial condition, which is of a Gaussian type with width controlled by the parameter $a$ and amplitude controlled by $\eta$.
\begin{eqnarray}
\label{a}
u(x, 0) &=& \frac{\eta}{\eta + (1 - \eta)\exp\left(\frac{x^2}{2a^2}\right)}  \\
h(x, 0) &=& \left[\frac{\eta}{1 - \eta}\right]\exp\left(-\frac{x^2}{2a^2}\right) 
\label{eta}
\end{eqnarray}
Then we took the convolution with the heat kernel with $h(x,0)$ and we get,
\begin{eqnarray}
h(x, t) = \frac{\eta a \exp\left[-\frac{x^2}{2(a^2 + 2t)}\right]}{(1 - \eta)\sqrt{a^2 + 2t}}
\end{eqnarray} 
From this $h(x,t)$ we get the exact solution as in the following, 
\begin{eqnarray}
u(x, t) = \frac{\eta a}{\left[\eta a + (1 - \eta)\sqrt{(a^2+2t)} \right] \exp\left\{-t + \frac{x^2}{2(a^2 + 2t)}\right\}}
\end{eqnarray}
The velocity of the saturation front can be obtained at the condition, $|x| << \sqrt{2t(a^2+2t)}$, where, the exponential term becomes constant. At this condition, we get, $ x=\pm \sqrt{2t(a^2+2t)} $ and thus we obtained the front velocity moving towards the right or along the $+x$ direction given by,
\begin{eqnarray}
\label{sv}
V(t)= \frac{a^2 + 4t}{\sqrt{2t \left( a^2 + 2t \right)}}
\end{eqnarray}
The asymptotic behaviour of the velocity $V$ can be calculated by expanding the above equation \eqref{sv} as a power series in as given by,
\begin{eqnarray}
V(t)=2 + \frac{a^4}{16 t^2} - \frac{a^6}{32 t^3} + \mathcal{O}\left(t^{-4}\right)
\end{eqnarray}
Then, the asymptotic velocity $V_{\infty}$ can be obtained by taking the limit $t\rightarrow\infty$ given by, 
\begin{eqnarray}
V_{\infty}=\lim_{{t \to \infty}} \frac{a^2 + 4t}{ \sqrt{2t \left( a^2 + 2t \right)}} = 2
\end{eqnarray}
Thus the asymptotic velocity $V_{\infty}$ falls to a constant value 2 in an exact manner as given by the above analysis. \\

{\noindent}We then follow another particular type of studies in \cite{ebert2,saarloos2} where through perturbation analysis technique, and numerical simulation of the PDE forward in time was done to track a particular front amplitude $(w)$. Since we now know the exact solutions, we can now directly evaluate any amplitude $(w)$ of the propagating wave, and a relevant one according to Fisher's technique \cite{fisher} is the one where we have half density or the mode of the heterozygote distribution. In order to achieve this condition, we first put $ u(x,t) = w =\frac{1}{2}$ and then we get a relation for $x$ and $t$ as given by,
\begin{eqnarray}
x_{w=\frac{1}{2}}= \sqrt{2(a^2 + 2t) \left[t+  \log\left(\frac{\eta a }{(1-\eta)\sqrt{a^2 + 2t}}\right)\right]}
\end{eqnarray}
From this we retrieve the results for the velocity corresponding to the amplitude $\omega=\frac{1}{2}$ of the wave,
\begin{eqnarray}
V_{w=\frac{1}{2}}=\frac{2 \log \left[\frac{\eta a }{(1-\eta)\sqrt{a^2 + 2t}} \right]+a^2 + 4t -1}{ \sqrt{2(a^2 + 2t) \left[ t+  \log\left(\frac{\eta a }{(1-\eta)\sqrt{a^2 + 2t}}\right) \right]}}
\end{eqnarray}
The asymptotic value of this velocity can be obtained by taking the limit $t\rightarrow\infty$, and we get,
\begin{eqnarray}
\lim_{{t \to \infty}} V_{w=\frac{1}{2}}=2
\end{eqnarray}
Similarly, we have done calculation for $w=\frac{1}{4}=0.25$ and $w=\frac{3}{4} =0.75$ respectively. The detailed calculations and expressions obtained can be found in the Appendix C. Next, we study these velocities ($V_w$) in a detailed manner and found the following observations.\\

\begin{figure}
    \centering
    \includegraphics[scale = 0.14]{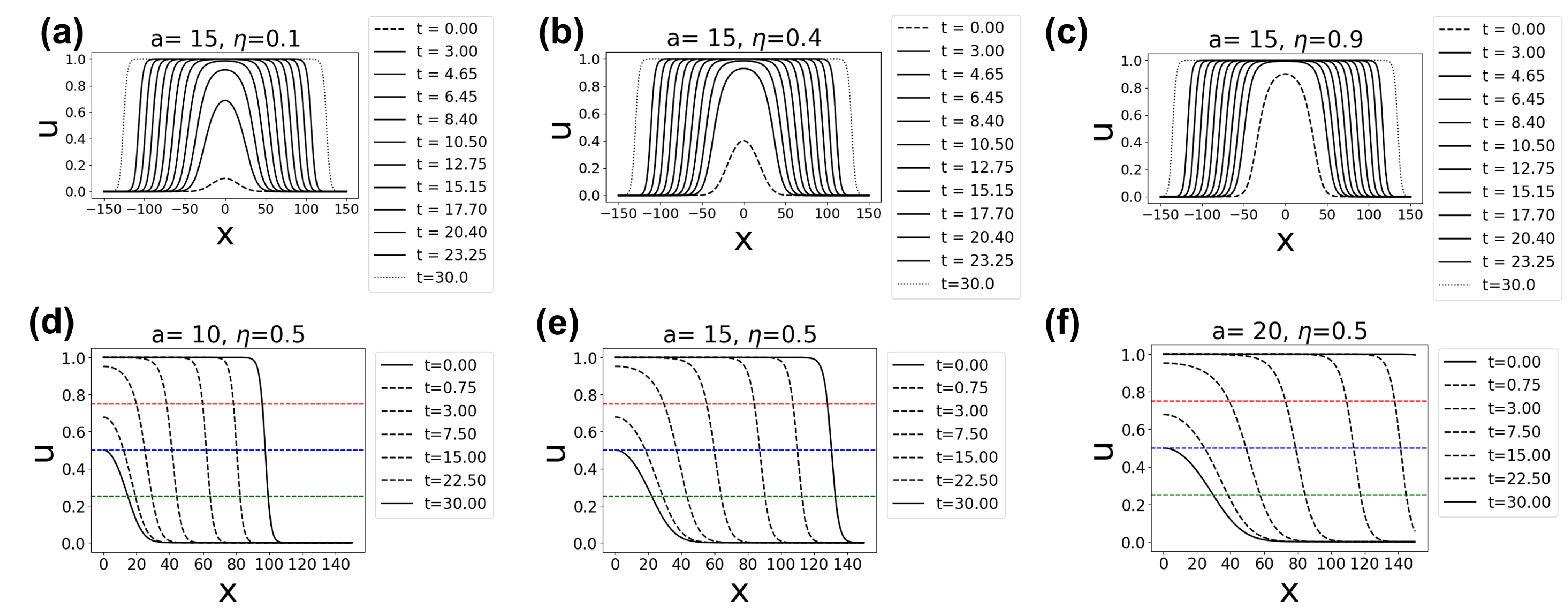}
    \caption{Exact solutions pertaining to the Gaussian type of initial condition with the width control parameter $a$ and amplitude control parameter $\eta$ for $F(u) = u(1-u)$. Waves progression with with $a=15$ with (a)$\eta =0.1$ (b) $\eta=0.4$ (c) $\eta=0.9$. This shows how the initial gradients effect the wave shapes transients. Now changing the width with a fixed amplitude $\eta = 0.5 $ with (d) $a=10$ (e) $a=15$ (f) $a=20$. Higher the width of the initial profile higher is the velocity of propagation temporarily. The three horizontal lines are drawn at amplitudes $w = 0.75,0.5,0.25$. The red line denotes w = 0.75 , blue for w=0.5, and green for w = 0.25. These are the velocities($V_w$) we track analytically. We show this graphically to see a clear indication of the dynamics of the shape of the front.}
    \label{fig3}
\end{figure}

{\noindent}First, we compare the effects of the height $\eta $ and the width $a$ on the travelling wave profile of the initial condition of the Gaussian distribution we took as given in \eqref{a} and \eqref{eta}. This has been graphically represented in the Fig.\ref{fig3}(a-f). The plots \ref{fig3}(a-c) show how the initial density $\eta$ effects the traveling wave form. Fig. \ref{fig3}(d-f) shows how the initial spread effects the transients. From the results shown in the figures it is evident that in the early transients we can see that $ u(x,t)$ shoots up in very short span of time probably due to the exponential growth term used in the transformation ($f(t)=e^t$), then attains saturation in the $u$ distribution with respect to $x$ for various various time slices, and then once the saturation develops the front of the wave starts translating forward in space. Further, the effect of changing the $\eta$ modulates the shape of the transient fronts as seen in Fig. \ref{fig3}(a-c). The effect of width parameter can be seen in Fig. \ref{fig3}(d-f), where, we can see that the higher the initial width greater is the time taken by the transients to relax to the asymptotic shape.\\

\begin{figure}
    \centering
    \includegraphics[scale = 0.14]{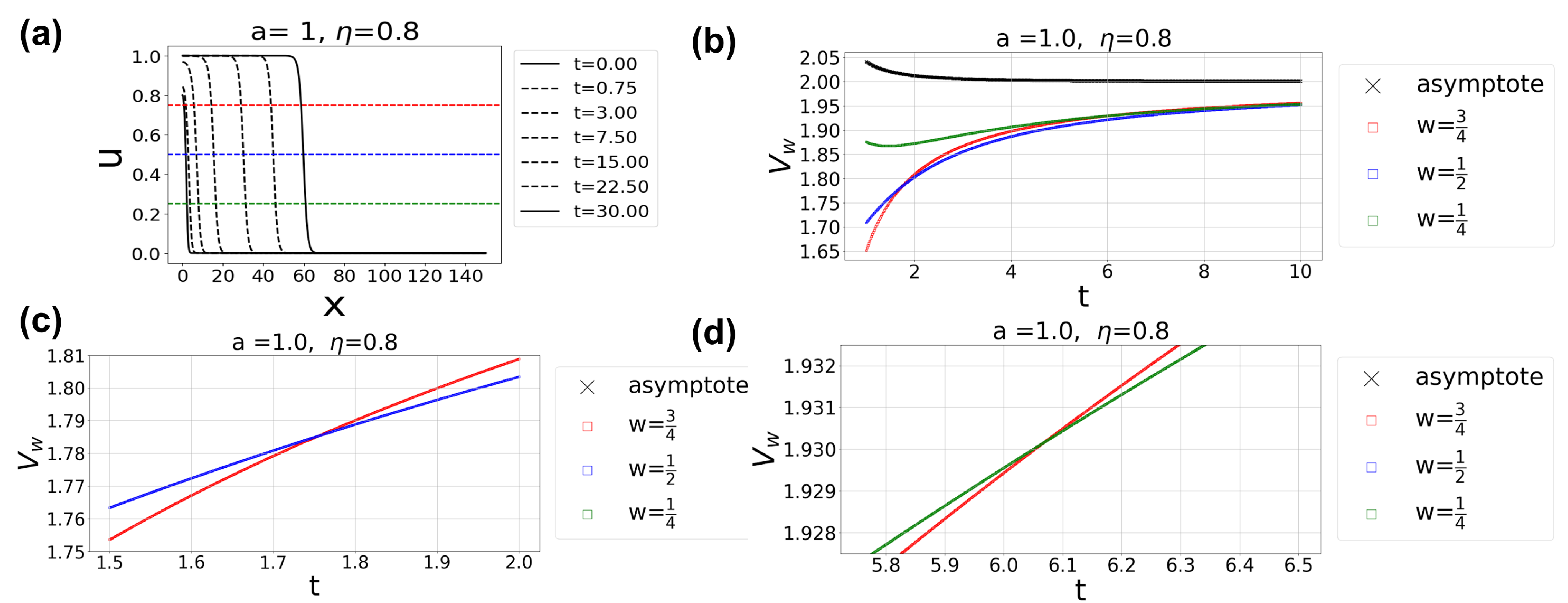}
    \caption{For a sufficient localized initial conditions with high initial density we can see the rapid relaxation of the velocities and the front shape. (a)The evolution showing the amplitudes(w) for which we track $V_w$. (b) The red line denotes w = 0.75 , blue w=0.5, and green w=0.25 amplitudes. The velocities can be seen crossing over each other as the gradient of the corresponding front amplitude changes with the front progressing ahead. (c) Velocity crossig over ,the red line (for w=0.75) crosses the blue one (for w=0.5) . (d) the subsequent crossing over of w=0.25. These rapid relaxations begs the question can these be observed in real experiments. }
    \label{fig4}
\end{figure}

{\noindent}We then study the wave propagation from a highly localized initial condition with $a = 1$ and $\eta =0.8$ as shown in Fig. \ref{fig4} to demonstrate the crossing over scenario as given in an intuitive picture in (Fig.3(a) of \cite{saarloos2} p. 6370). We found that the front shape changes significantly with respect to wave amplitude $w$ by studying the relaxation of the front velocity $V_w$ as a function of $w$. The velocity of $w=0.75$ (red line) relaxes rapidly and crosses over the velocities of $w=0.25$ (green) and $w=0.5$ (blue line) as shown in the Fig. \ref{fig4}(c)-(d). This verifies a beautiful result that the the higher value of the front has a higher velocity but is located in space behind the lower value which has a lower velocity. We also point out that all these velocities are relaxing asymptotically to the value 2 from below. For practical purposes the crossover regimes cannot be resolved for such highly localized initial conditions.\\

\begin{figure}
    \centering
    \includegraphics[scale = 0.14]{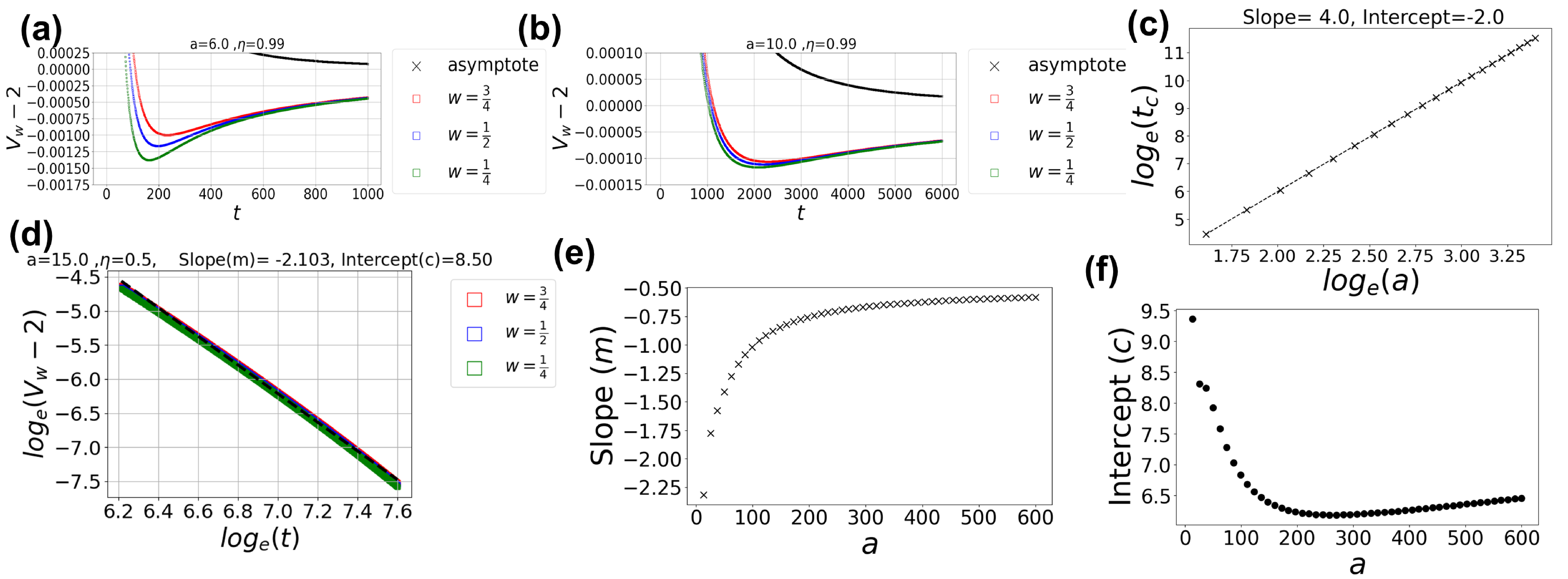}
    \caption{The initial regime of the transient relaxation of the front velocities. We study in detail how the initial velocities relax to the lower values of velocities. (a)-(b) Two distinct regimes can be observed where in the initial regime it falls below 2 and in the second regime it approaches 2 from below. i.e. $V_w-2$ changes sign. (c) Crossover times for various values of parameter $a$ from one relaxation regime to another for $V_{w=0.5}$ for $\eta=0.5$. (d) The relaxation can be written as the form $V_w = 2 + \frac{e^{c(a)}}{t^{m(a)}}$ with the slope (m) and the intercept (c) as a function of the parameter $a$ as given from the log-log graph.(e) The slope(m) vs $a$ shows the exponent of t showing a trend that for larger values of $a$ we get a $\frac{1}{\sqrt{t}}$ dependence. (f) The intercept(c) as a function of the initial condition spread parameter $a$. }
    \label{fig5}
\end{figure}

\begin{figure}
    \centering
    \includegraphics[scale = 0.14]{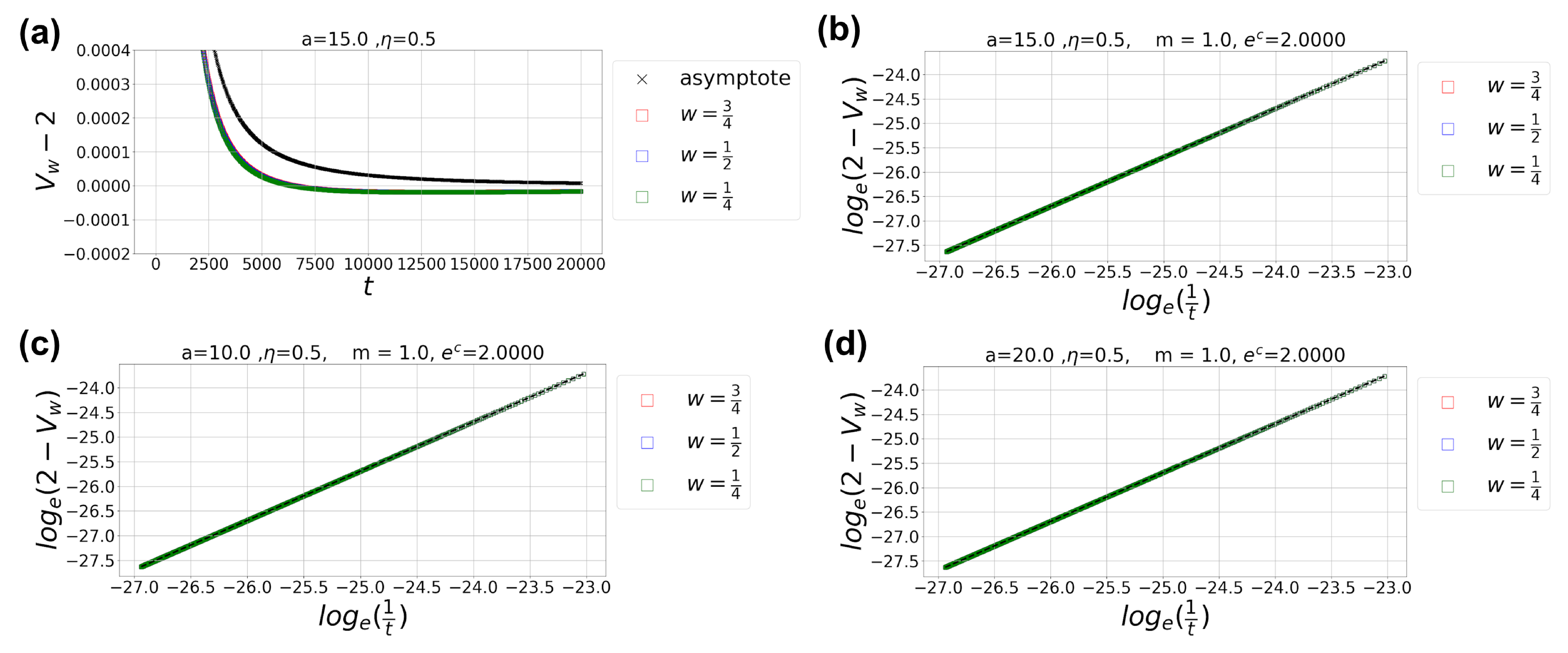}
    \caption{ (a) The relaxation of the front velocity($V_w$) for larger times for illustrating how the velocities relax asymptotically. (b-d) The intial condition independent relaxation of velocities asymptotically with time can be traced through a relation as ($V_w = 2-\frac{m}{e^c t} =2-\frac{1}{2t}$).}
    \label{fig6}
\end{figure}

{\noindent}Now, we focus on the study of long and short times relaxation of the velocities ($V_w$). The results are shown in details in the plots in Fig. \ref{fig5} and \ref{fig6}. We clearly found from the plots that there are two distinct regimes of relaxation of the front velocities. The first regime depicted by the short times behaviour, we found the velocities fall to the asymptotic velocity from above i.e. $ V_w-2 >0$, and, then in the second regime for long times, we observed the velocities approach to asymptotic value from below i.e. $ V_w-2 <0$.\\

{\noindent}Next, to study the dynamics of the crossover, we took a certain range of the initial conditions and then check how does this crossover ($ V_w-2 <0$) is obtained and it's progress with time into the second regime of relaxation of the front velocities. We then estimated the point at which $ V_w-2 $ changes sign as seen in the plots Fig.\ref{fig5}(a,b). This transient behaviour was also predicted in \cite{saarloos3}. The crossing over epoch $t_c$ can be estimated as follows,
\begin{eqnarray}
\label{tc}
t_c = \frac{a^4}{e^2}
\end{eqnarray}
We also see from the Fig.\ref{fig5}(c) that if the initial conditions are not tightly localized we would observe very high velocities of propagation for long period of times before we can actually observe the slowing down of the front of the propagating wave. This suggests that if the mutant dispersal is positively correlated with the density then the waves might be temporarily accelerated due to the gradient effects in the habitat landscape. This point was also raised by Fisher in (Para 1 of Section IV in \cite{fisher} p.359).\\

{\noindent}We study in detail the nature of the propagating wave near the crossover as shown in Fig. \ref{fig5}(d). The results indicate that during this relaxation regime, the velocities fall to the asymptote velocity from above i.e. $V_w-2 >0$ and thus this velocity is given by, 
\begin{eqnarray}
\label{asv}
V_w= 2 + \frac{e^{c(a)}}{t^{m(a)}}
\end{eqnarray} 
where $m(a)$ and $c(a)$ are the fitted slopes and intercepts of the log-log plots as shown in \ref{fig5}(d). From this results, we first observe that the $m(a)$ in the plot \ref{fig5}(e) exhibit nonlinear behaviour as a function of $a$ for smaller values of $a$, where, we found that the fall of the initial transient velocities is roughly $\displaystyle\lim_{a\rightarrow small}V_{w}\sim O\left (\frac{1}{t^2}\right )$. Then, as $a$ increases, $m$ saturates towards the value $\displaystyle\lim_{a\rightarrow\infty}m(a)=-0.5$, for higher values of $a$, where, the fall of transient velocity is dominated roughly by $\displaystyle\lim_{a\rightarrow large}V_{w}\sim O\left(\frac{1}{\sqrt{t}}\right)$. Secondly, we further observe the behaviour of intercept $c(a)$ with $a$ quite differently, where, $c(a)$ monotonically decreased with $a$ till a minimum value of $c$ i.e. $c_{min}$ as shown in the Fig.\ref{fig5}(f), then $c(a)$ increases slowly as a function of $a$ indicating nonlinear nature of $c(a)$ with respect to $a$.\\

{\noindent}On the other hand, for the second regime of $V_w$ ($V_w-2< 0$) approaching the asymptotic velocity from the below and this relaxation can be given by, 
\begin{eqnarray}
V_w=2-\frac{1}{2t}
\end{eqnarray} 
independent of the initial conditions as seen in the plots in Fig. \ref{fig6}(f,g,h).  Hence, through these exact solutions we have thus re-verified the conjectures regarding the relaxation of front velocities in \cite{fineberg,ahlers,saarloos2} . Specially we would like to point out similar results in (section V of \cite{powell}) and (Fig.1 in \cite{ebert2}p.1652.), and  Niklas et.al. (Fig2. of \cite{krumbhaar} p.40) and in Kockelkoren et.al. (Fig. 1 in \cite{saarloos3} p.171 ) where such type of results were obtained using numerical integrations of various similar classes of PDE. \\

\begin{figure}
    \centering
    \includegraphics[scale = 0.14]{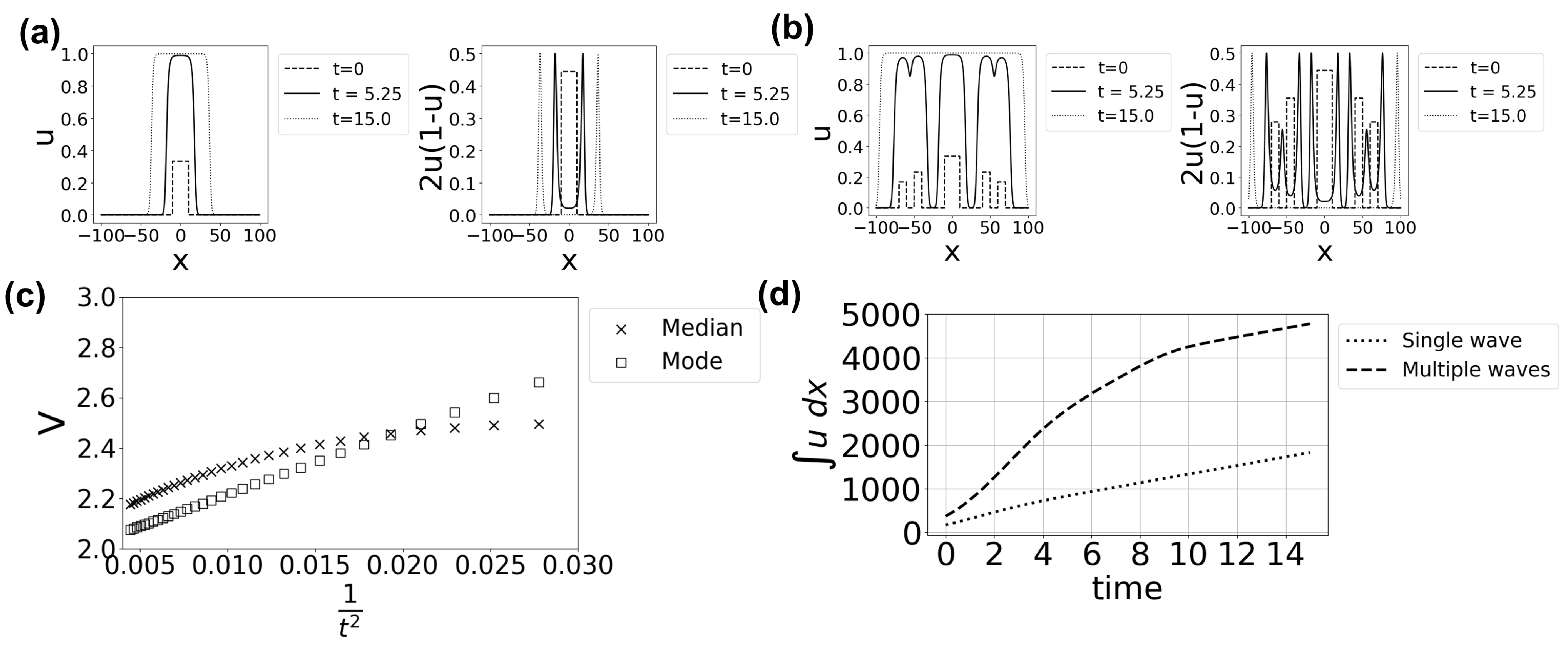}
    \caption{(a)Evolution for the step function type of initial conditions. (b)Multiple locations initial conditions where the mutant arise as a localized type of initial conditions. (c) The velocity of the front can be tracked by using the median and the mode of the heterozygote distribution. The mode shows excellent agreement with a $ \frac{1}{t^2}$ relaxation.(d) Verifying the hypothesis that multiple waves propagating in the habitat at the same time leads to wider area coverage. }
    \label{fig7}
\end{figure}

{\noindent}We now proceed to verify another important contextual conjecture raised by Fisher in \cite{fisher}. This calculation proves how the transformation we have used can be effectively utilized to gain deep theoretical insights into how the mutant species can spread far away from its origin in the habitat landscape. We assume pockets of mutant arise at many locations with a step function type initial conditions as shown in the Fig.\ref{fig7}. As an example this scenario may be given as an initial condition of the type below:

\begin{eqnarray}
u(x,0) =
\begin{cases}
    0.16 & \text{if } -70 \leq x \leq -60 \\
    0.23 & \text{if } -50 \leq x \leq -40 \\
    0.33 & \text{if } -10 \leq x \leq 10 \\
    0.23 & \text{if } 40 \leq x \leq 50 \\
    0.16 & \text{if } 60 \leq x \leq 70 \\
    0 & \text{otherwise}
\end{cases}
\end{eqnarray}
which gives $h(x,0) = \frac{u(x,0)}{1-u(x,0)}$, and then the solution from convolution with the heat kernel given by, 
\begin{eqnarray}
h(x,t) &=& a_1+a_2+a_3+a_4+a_5\\
&&a_1= 0.1 \left[\text{erf} \left( \frac{-60 - x}{2 \sqrt{t}} \right) + \text{erf} \left( \frac{70 + x}{2 \sqrt{t}} \right) \right]\nonumber\\ 
&&a_2= 0.15 \left[ \text{erf} \left( \frac{-40 - x}{2 \sqrt{t}} \right) + \text{erf} \left( \frac{50 + x}{2 \sqrt{t}} \right) \right]\nonumber\\
&&a_3= 0.25 \left[ \text{erf} \left( \frac{10 - x}{2 \sqrt{t}} \right) + \text{erf} \left( \frac{10 + x}{2 \sqrt{t}} \right) \right]\nonumber\\
&&a_4= 0.15 \left[\text{erf} \left( \frac{50-x}{2 \sqrt{t}} \right) - \text{erf} \left( \frac{40-x}{2 \sqrt{t}} \right) \right]\nonumber\\
&&a_5= 0.1 \left[\text{erf} \left( \frac{70-x}{2 \sqrt{t}} \right) - \text{erf} \left( \frac{60-x}{2 \sqrt{t}} \right) \right]\nonumber
\end{eqnarray}
and finally we obtain,
\begin{eqnarray}
\label{uxt}
u(x,t) = \frac{e^t h(x,t)}{1+e^t h(x,t)}
\end{eqnarray}
We now plot the solution using the above equation \eqref{uxt} as shown in the Fig.\ref{fig7}(d) using the same discretization method. The results verify an important conjecture proposed by Fisher in (\cite{fisher} Section VI p.367 ) which explained the possibility of multiple such waves as shown in Fig. \ref{fig7}(d) can help the fixation of the advantageous mutant in a wide area than the possibility of fixation by a single wave. \\

{\noindent}Next, we further study by taking into account the biallelic case. We considered the parent allelomorph as $A$ and the mutant as $a$. The frequencies, relative fitness of these parameters in this model are given in Table 1. To each of the cases, the corresponding change in $p$ in one generation i.e. $\Delta p = F(p)$ terms are given in the Table 1.\\

\begin{table*}[htbp]
  \centering
   \begin{tabular}{p{3.9cm}p{2.5cm}p{2.5cm}p{2.5cm}p{4cm}}

      Genotypes: & AA & Aa & aa & \\
      \hline
     Frequencies: & $q^2$  & $2pq$ & $p^2$ &\\
     \hline
     Relative fitnesses: &&&& Non-linear term $kF(p)$\\
     \hline
    Case I & 1 & 1-k & 1 &  $ F(u) =u(1-u)(2u-1)$ \\
    \hline
    Case II & 1-k & 1 & 1-k &  $ F(u) =u(1-u)(1-2u)$ \\
    \hline
  \end{tabular}
\caption{Table of various cases on the fitness condition of the mutant .}
\label{t1}
\end{table*}
{\noindent}The fixed points of $F(p)$ are the points $p^*$ for which $F(p^*)=0$ .The stability of $p^*$ are the determining criteria for the existence of wave type solutions. The various model cases given in Table \ref{t1} are derived using Eq.5.2.13(p.182) and Eq.5.3.6 (p. 192 ) of \cite{crow}. In case-I, we consider a case that the heterozygote is less fit than the homozygotes. However, the case-II pertains to situation, where, the heterozygote has more fitness. This situation is of special importance because an additional fixed point of F(p) within the range $0\leq p \leq 1$ can be obtained in such situation playing crucial role in the wave propagation. The case-I describes a threshold effect and case-II gives a hairtrigger effect as described in the section 6 and section 3 of the work \cite{aronson2} respectively. These systems also show the delicate nature of the the polymorphisms that can arise in certain genetic systems. 

{\noindent}We now substitute the functional forms of $F$ as given in Table 1 to the equation \eqref{meq}, and then apply the Proposition 1 for further analysis. We considered and calculated for the two cases mentioned in Table 1 as given below.\\

\noindent\textbf{Case I: $F(u)=u(1-u)(2u-1)$}\\

{\noindent}This system has three fixed points given by, $(0,\frac{1}{2},1)$, which are found to be stable, unstable, stable respectively. Since the case-I in Table 1 gives the functional form of $F$ as,
\begin{eqnarray}
F(u)=u(1-u)(2 u-1)=3 u^{2}-u-2 u^{3}
\end{eqnarray}
we have its derivative, $F^{\prime}(u)=6 u-1-6 u^{2}$, and, hence, it shows that $uF'(0) \not\geqslant F(u), ~ \forall u \in[0,1]$. Substituting these $F(u)$ and $F^\prime(u)$ to the PDE \eqref{meq}, the equation becomes,
\begin{eqnarray}
\label{c1}
u_{t}=\left\{u_{x x}+u_{x}^{2}\left[\frac{\frac{f'(t)}{f(t)}-6 u+1+6 u^{2}}{u(1-u)(2 u-1)}\right]\right\} +u(1-u)(2 u-1)
\end{eqnarray}
From this equation \eqref{c1} and using the functions $F(u)$ and $F^\prime(u)$ to the equation \eqref{trans}, we get,
\begin{eqnarray}
G(u)= f(t)h(x,t)=exp\left[ \frac{f'(t)}{f(t)} \int \frac{du}{F(u)}\right] = \left[ \frac{(1-2 u)^{2}}{(1-u) u}\right] ^{\frac{f'(t)}{f(t)}} 
\end{eqnarray}
 From this calculations, we get that the inverse transformation $G^{-1}$ has splitted our initial conditions into two sets as given by,
 \begin{eqnarray}
 R1:=0<u(x,0)<\frac{1}{2}\quad  \text{or} \quad R2:=\frac{1}{2}<u(x,0)<1\nonumber
 \end{eqnarray}
Hence, we have the following results :
\begin{enumerate}
\item\textbf{R1: $\forall u(x,0) \in  0<u(x,0)<\frac{1}{2}  $} \\
{\nonumber}The solution of the equation \eqref{c1} for the initial condition given by $R1$ converges to the fixed point 0 asymptotically. Hence, we get the solution $u_{-}$ for $R1$ as,
\begin{eqnarray}
u_{-}(x,t) = \frac{1}{2} - \frac{1}{2} \sqrt{\frac{[f(t)h(x,t)]^{\frac{f(t)}{f'(t)}}}{ 4+[f(t)h(x,t)]^{\frac{f(t)}{f'(t)}} }}\nonumber
\end{eqnarray}

\item\textbf{R2: $\forall u(x,0)\in  \frac{1}{2}<u(x,0)<1 $}\\ 
{\noindent}The solution of the equation \eqref{c1} for the initial condition given by $R2$ converges to the fixed point 1 asymptotically. Hence we get the solution $u_{+}$ for $R2$ as,  
\begin{eqnarray}
u_{+}(x,t) = \frac{1}{2} + \frac{1}{2} \sqrt{\frac{[f(t)h(x,t)]^{\frac{f(t)}{f'(t)}}}{ 4+[f(t)h(x,t)]^{\frac{f(t)}{f'(t)}} }}\nonumber
\end{eqnarray}
\end{enumerate}

{\noindent}Thus, by using our proposed method, we could able to retrieve the same result as was proven in Proposition 6.1 of \cite{aronson2}. Further, these results also verify our assumption taken while assigning the fitness for the cases discussed above indicating that the heterozygotes (Aa) have a fitness disadvantage relative to the homozygotes (aa or AA). Hence, depending on its initial conditions either the mutant survives or it is wiped out.\\  

{\noindent}Now, let us take an application, where, we first take an initial condition in $R2$ region and $f(t)=e^t$, such that, $\frac{f(t)}{f'(t)} =1$. This initial condition is such that  the mutant population is saturated up to $b$ units length behind the origin and falls after $b$ units in front of the origin to adjust above half density. Such an initial condition is given by,
$u(x,0)= \frac{1}{2} +\frac{1}{2} \sqrt{ \frac{e^{-b x}}{4+e^{-b x}}}$, such that, $u(x,0) \in R2$ region. Following the above procedure, this initial condition gives us $ h(x,0) = e^{-b x}$ from which we get, $h(x,t) = e^{-b (x- b t)} $, and thus, we obtain the following closed form solution,
\begin{eqnarray}
\label{wvf}
u(x,t) = \frac{1}{2} +\frac{1}{2} \sqrt{ \frac{1}{1+4 e^{b [x- (b+\frac{1}{b}) t]}}}
\end{eqnarray}
Using this equation \eqref{wvf} gives us for the velocity of the saturation front $V=b + \frac{1}{b}$. \\

\begin{figure}
    \centering
    \includegraphics[scale = 0.16]{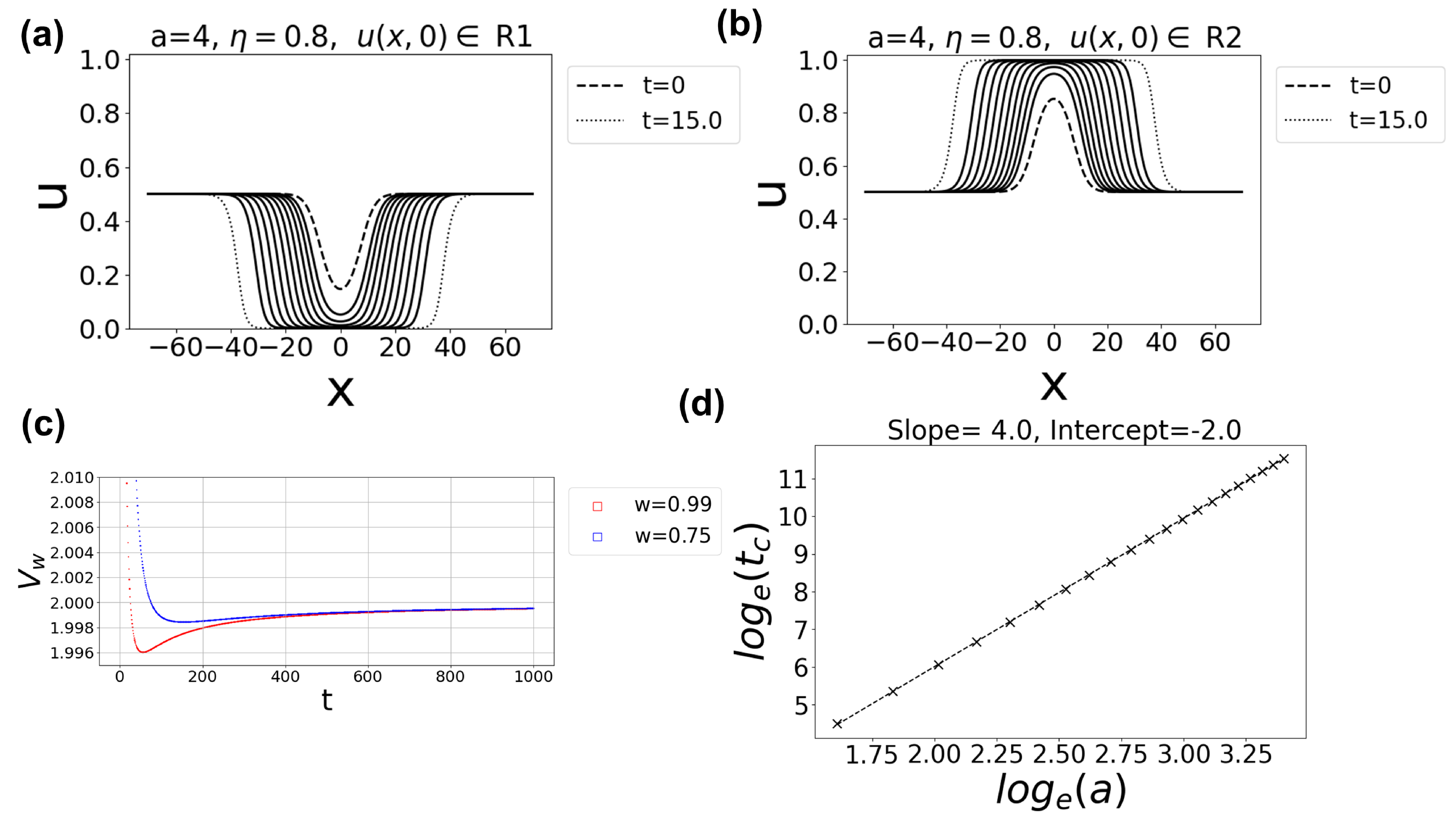}
    \caption{The heterozygote inferior fitness case showing the threshold effect. (a)The initial condition has to be above the threshold value to grow to a steady state value of 1. (b)Any gradient below the threshold will lead to asymptotic fall to the 0 steady state value of the mutant density. (c) The relaxation of the front velocities.(red for w=0.99, blue for w=0.75) (d) The crossing over times ($t_c$) for the velocity($V_w$) of the front amplitude value $w=0.75$.  }
    \label{fig8}
\end{figure}

{\noindent}We then take a symmetric distribution about the origin as is given by Gaussian distribution which is again in region $R2$, where, it raises around the origin about a fraction $\eta$ and is spread for a width $a$ as shown in the Fig. \ref{fig8}(a). This is given by, $u(x,0)= \frac{1}{2} +\frac{1}{2} \sqrt{ \frac{\frac{\eta}{1-\eta} \exp\left(\frac{-x^2}{2a^2}\right)}{4+\frac{\eta}{1-\eta}\exp\left(\frac{-x^2}{2a^2}\right)}}$. This gives us the corresponding function as $h(x,0) = \frac{\eta}{1-\eta} e^{\frac{-x^2}{2a^2}} $, and, thus we have, $h(x,t) = \frac{\frac{\eta a}{1-\eta} \exp\left[-\frac{x^2}{2(a^2 + 2t)}\right]}{\sqrt{a^2 + 2t}}$. Using this function $h(x,t)$, we arrive at the exact solution given by,
\begin{eqnarray}
\label{uc1}
u(x,t) = \frac{1}{2}+\frac{1}{2}\sqrt{ \frac{e^{t}\left[\frac{\eta ae^{-\frac{x^{2}}{2a^{2}+4t}}}{(1-\eta)\sqrt{a^{2}+2t}}\right]}{4+e^{t}\left[\frac{\eta ae^{-\frac{x^{2}}{2a^{2}+4t}}}{(1-\eta)\sqrt{a^{2}+2t}}\right]}}
\end{eqnarray}
Now, when $|x| << \sqrt{2(a^2+2t)}$ the exponential term in the equation \eqref{uc1} becomes $e^{-\frac{x^{2}}{2a^{2}+4t}}\rightarrow 1$, and we have the asymptotic saturation behaviour of $u(x,t)$. Thus we get the asymptotic velocity again as before $V= \frac{a^2 + 4t}{\sqrt{2t \left( a^2 + 2t \right)}}$. 
Similarly, we get the relation for $u(x,t) = \frac{3}{4}$ between $x$ and $t$ as,
\begin{eqnarray} 
x&=& \sqrt{2(a^2 + 2t)\left[ t +  \ln\left( \frac{3\eta a}{4(1-\eta) \sqrt{a^2 + 2t}} \right)\right]}\\
V_{w=\frac{3}{4}}&=& \frac{2 \log \left[\frac{ 3 \eta a}{4(1- \eta ) \sqrt{a^2+2 t}}\right]+a^2+4 t-1}{ \sqrt{2\left(a^2+2 t\right) \left[\log \left(\frac{  3\eta a}{4(1-\eta ) \sqrt{a^2+2 t}}\right)+t\right]}}\nonumber
\end{eqnarray}
Hence, the large $t$ asymptotic value of the velocity is constant, $\lim_{{t \to \infty}} V_{w=\frac{3}{4}}=2$.\\

\noindent The threshold effect is already built intrinsically into the transformation as it had split our initial conditions into two regions. For initial conditions above the particular value the steady state 1 is reached asymptotically while for the values below state 0 is reached asymptotically. We have plotted some graphical representations of the exact solutions and we can see how the effect propagates. In the first Fig.\ref{fig8}(a), the initial condition is well above the threshold value hence it reaches the state 1 asymptotically. Similarly in the second Fig.\ref{fig8}(b) we see how the heterozygotes are wiped out if there is a small downward gradient in the density at a particular location. The velocity relaxation observed in the Fig.\ref{fig8}(c) again follows the similar behaviour with a crossover and then approaching the asymptotic value from below. We have shown Fig.\ref{fig8}(d) the cross over time ($t_c$) with respect to the $V_{w=0.75}$. We recover the similar result as from before.\\

\noindent\textbf{Case II: $F(u)=u(1-u)(1-2u)$}\\

\noindent  This system corresponds to the case where the heterozygote has an advantage over the homozygotes in terms of fitness. This also occurs if the gene becomes less beneficial as its frequency increases (p.7 of \cite{crow1} ). The system has 3 fixed points $0,\frac{1}{2},1$ which are unstable, stable and unstable respectively. Now, proceeding in the same way as we did in previous cases we have,
\begin{eqnarray}
\label{fc2}
F(u)&=&u(1-u)(1-2 u)=u-3 u^{2}+2 u^{3} \\  
F^{\prime}(u)&=&1-6 u+6 u^{2}
\label{fpc2}
\end{eqnarray}
Substituting these expressions, the PDE \eqref{meq} becomes,
\begin{eqnarray}
\label{c2}
u_{t}=\left\{u_{x x}+u_{x}^{2}\left[\frac{\frac{f'(t)}{f(t)}-(1-6 u+6 u^{2})}{u(1-u)(1-2u)}\right]\right\} +u(1-u)(1-2u)
\end{eqnarray}
Using the expressions \eqref{fc2} and \eqref{fpc2} to the equation \eqref{c2}, and then using the transformation \eqref{trans}, we have, 
\begin{eqnarray}
\label{gut}
G(u)= f(t)h(x,t)=exp\left[ \frac{f'(t)}{f(t)} \int \frac{du}{F(u)}\right] = \left[ \frac{(1-u) u}{(1-2 u)^{2}}\right] ^{\frac{f'(t)}{f(t)}} 
\end{eqnarray} 
From this equation \eqref{gut}, we get the inverse transformation function $G^{-1}$, and this transformation has splitted our initial conditions into two sets given by,
\begin{eqnarray}
R1:=0<u(x,0)<\frac{1}{2}\quad  or \quad R2:=\frac{1}{2}<u(x,0)<1
\end{eqnarray}
Hence, in our analysis, we deal only with $R1$ and $R2$ initial conditions, and we have the following results :

\begin{enumerate}
\item For the initial condition \textbf{R1: $\forall u(x,0) \in 0<u(x,0)<\frac{1}{2} $}, we get the following solution,
\begin{eqnarray}
u(x,t)=\frac{1}{2} - \frac{1}{2} \sqrt{\frac{1}{1+ 4[f(t)h(x,t)]^{\frac{f(t)}{f'(t)}}}}
\end{eqnarray}

\item For the initial condition \textbf{R2: $\forall u(x,0) \in \frac{1}{2}<u(x,0)<1 $}, we obtained the following solution,
\begin{eqnarray}
u(x,t)=\frac{1}{2} + \frac{1}{2} \sqrt{\frac{1}{1+ 4[f(t)h(x,t)]^{\frac{f(t)}{f'(t)}}}}
\end{eqnarray}
\end{enumerate}

{\noindent}This result again verifies the hair-trigger effect proved in Theorem 3.1 of \cite{aronson2}.\\

\begin{figure}
    \centering
    \includegraphics[scale = 0.153]{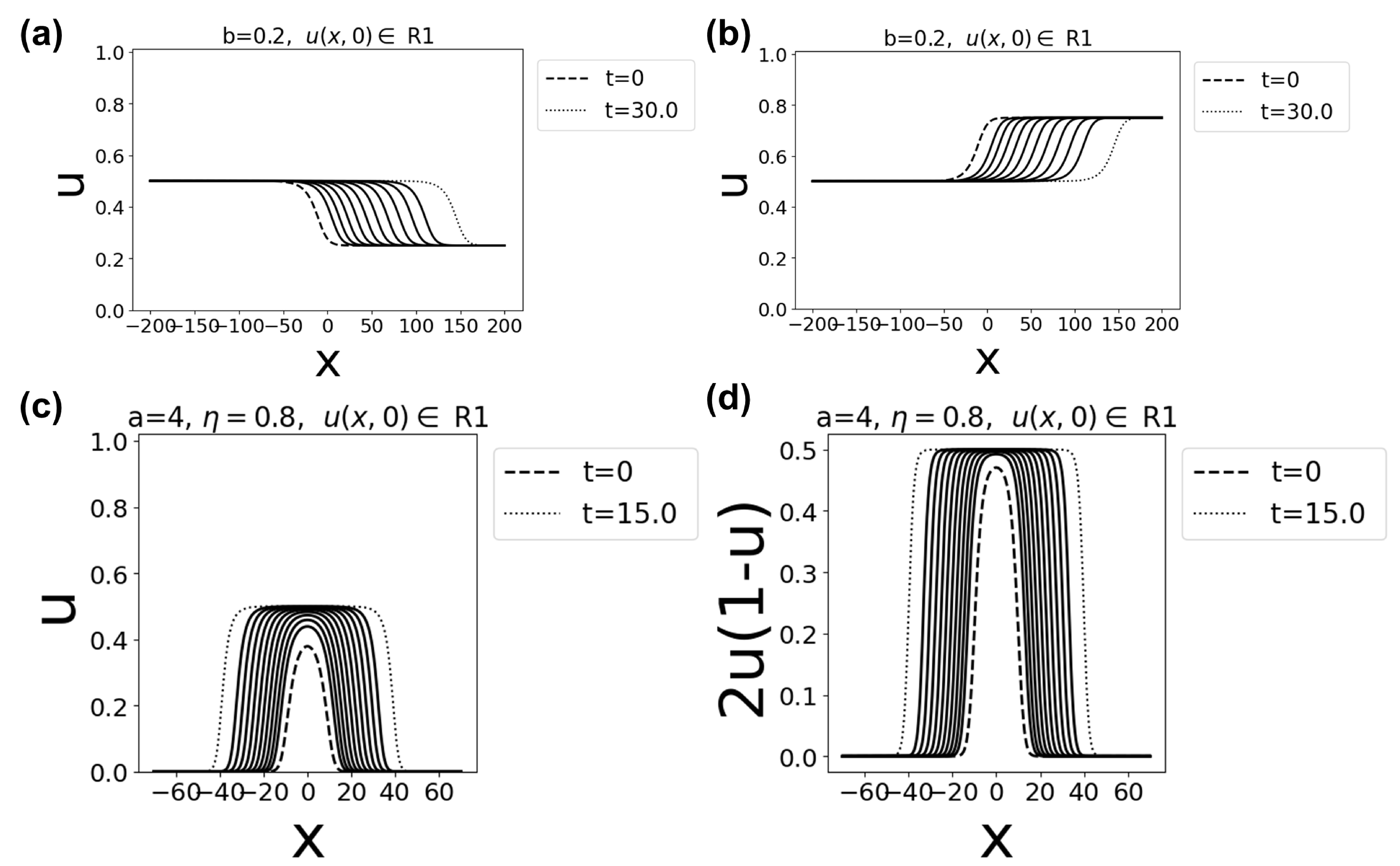}
    \caption{For the heterozygote superior fitness case showing the hair-trigger effect. This case was shown to see how we can have a stable polymorphism with all the types i.e. AA, Aa, aa co-existing in the habitat. (a)-(b) Progression of the mutant wave. We can see the mutant density decreases to 0.5 everywhere in the habitat with shape parameter b=0.2. The initial condition is a sigmoid shape. (c)-(d) The progression of the wave with the Gaussian type of initial conditions. The heterozygote distribution(d) can be seen expanding everywhere in the habitat.  }
    \label{fig9}
\end{figure}

{\noindent}Now, we chose the sigmoidal type initial condition as shown in \ref{fig9}(a-b) as this function is quite for natural growth processes \cite{Yin}. We first take an initial condition in $R2$ region and $f(t)=e^t$, such that, $\frac{f(t)}{f'(t)} =1$ . This initial condition is such that the mutant population is saturated upto $b$ units length behind the origin and falls after $b$ units in front of the origin just above the half density. Then, the initial condition of $u$ satisfying such situation is of the form, $u(x,0)= \frac{1}{2} +\frac{1}{2} \sqrt{ \frac{1}{4+e^{-b x}}}$, such that, $u(x,0) \in R2$ region. This gives $ h(x,0) = e^{-b x}$, from which we can obtain $h(x,t) = e^{-b (x- b t)} $. Using these functions we arrive at the following a closed form, 
\begin{eqnarray}
u(x,t) = \frac{1}{2}+\frac{1}{2}\sqrt{\frac{1}{4+e^{-b\left[x-\left(b+\frac{1}{b}\right)t\right]}}}
\end{eqnarray}
On the other hand, if we take the Gaussian type of initial condition and the same functional form $f(t)=e^t$, we have the inial condition in $u$, 
$\displaystyle u(x,0) = \frac{1}{2}+\frac{1}{2}\sqrt{\frac{1}{1+\frac{4\eta}{1-\eta}e^{\frac{-x^{2}}{2a^{2}}}}}$, from which we get,
\begin{eqnarray}
u(x,t) = \frac{1}{2}+\frac{1}{2}\sqrt{\frac{1}{1+\frac{4cae^{t}}{1-c}\frac{e^{-\frac{1}{2}\frac{x^{2}}{a^{2}+2t}}}{\sqrt{a^{2}+2t}}}}
\end{eqnarray}
For this case, the results of the distribution of $u(x,t)$ with $x$ for different time slices are shown in Fig. \ref{fig9}(c-d). As evident from the results in the figure, it can be seen that this evolution leads to the maintenance of the heterozygote population everywhere. The velocity relaxation for this case will be the same as before. \\

\noindent\textbf{Case III} $F(u) = u(1-u)(1+mu)$ \\

\begin{figure}
    \centering
    \includegraphics[scale = 0.54]{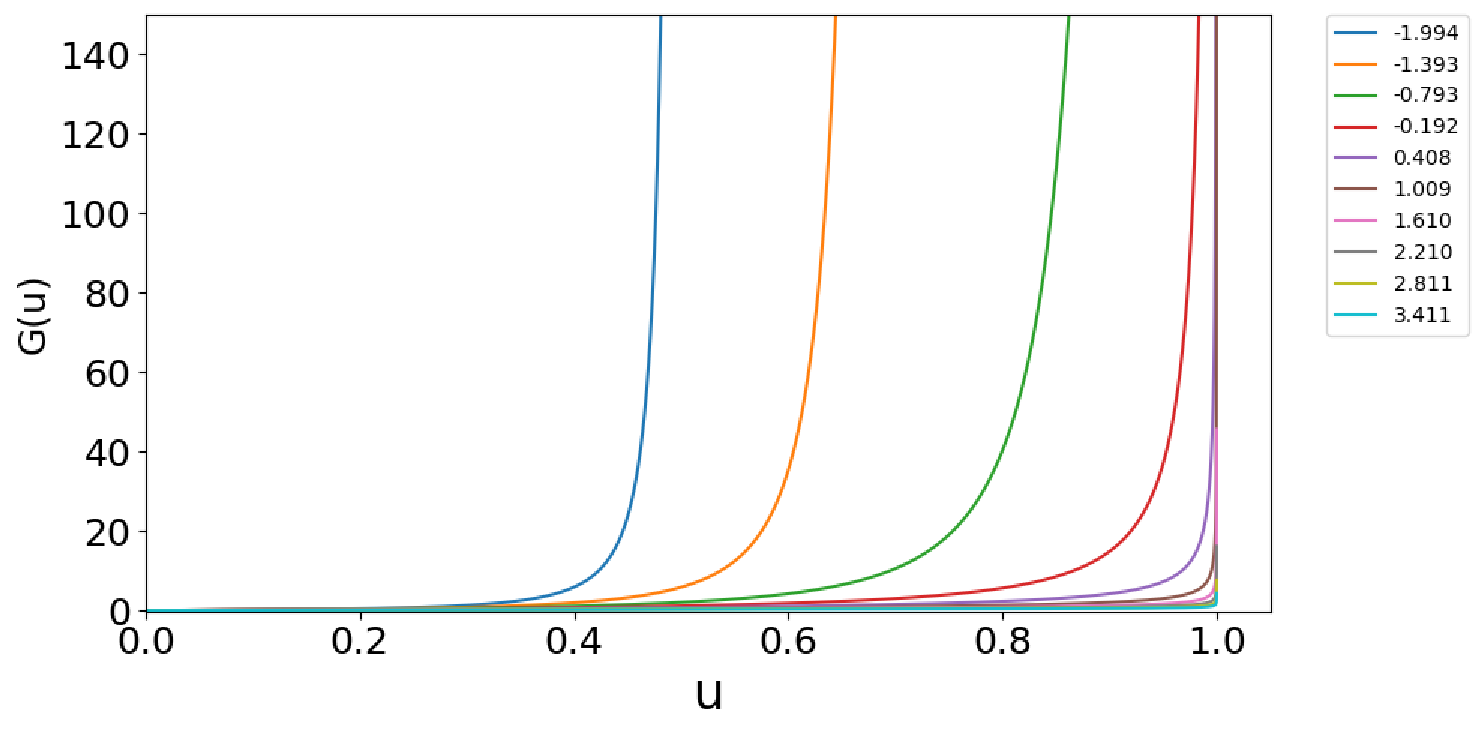}
    \caption{The transformation function G for various forms of the nonlinearity $F(u) = u(1-u)(1+mu)$ . The nature of the asymptotic increase of the G can be seen changing with m. The asympote increase shows the value(u) of the fixed point which is stable. }
    \label{fig10}
\end{figure}

{\noindent}Now, we consider a certain generalised form of the functional form $F$ considered in the case-II by defining a constant $m$ in the function. It is well established that this nonlinearity driven by $m$ shows a transition at a critical value of $m$ represented by, $m_c=2$ between pulled and pushed cases driven by the parameter $m$. We generated the results of the transformed function $G$ as a function of $u$ for some values of $m$ and check how the nonlinearity effects the transformation functions as shown in the Fig. \ref{fig10}. The plots of the results in the Fig. \ref{fig10} shows how the transformation function $G$ changes as $m$ is varied indicating how $G^{-1} $ will help us obtain saturation value. 

\subsection{Transformation ansatz type-II}

{\noindent}In this section, we propose a certain transformation of type $G(u)= f(t)h(x,t)$ that can linearize the Fisher-KPP PDE \eqref{fkpp}, where the function $h(x,t)$ does not follow a homogeneous diffusion equation but another PDE which is in a completely solvable class. This class of ansatz is similar in spirit to the matched asymptotics like solution class, where, the phenomena dictates what kind of solution spaces are allowed \cite{needham1,needham2}. For this problem, we have found particular cases that can completely linearize the original PDE in terms of solutions of known solvable PDE. They are as follows :

\subsubsection{Case A: $ f(t)= e^t$ and $h(x,t)$ satisfies  $h_t = h_{xx} - h $ }

{\noindent}We impose $h(x,t)$ to satisfy a linear diffusion equation of the form, 
\begin{eqnarray}
\label{difh}
h_t = h_{xx} - h
\end{eqnarray} 
We begin with the following transformation,
\begin{eqnarray}
G(u)&=& f(t)h(x,t)\nonumber\\ 
h_t &=& \frac{G'u_t}{f} - \frac{f'}{f}h \nonumber\\
h_{xx} &=& \frac{G''(u_x)^2}{f}  + \frac{G'u_{xx}}{f} 
\end{eqnarray}
This transformation is now put in the equation \eqref{difh}. Rearranging the terms, we obtain the following transformed equation in $u$,
\begin{eqnarray}
u_t +\left(1- \frac{f'}{f}\right) \frac{G}{G'}  = \frac{G''}{G'} u_x^2 + u_{xx}
\end{eqnarray}
Now, comparing with \eqref{fkpp} we get the transformation $G$ as, 
\begin{eqnarray}
G(u(x,t))=\exp \left[\int \frac{du}{F(u)}  \right] 
\end{eqnarray}
which gives the following PDE 
\begin{eqnarray}
\label{tru}
u_t +\left(1- \frac{f'}{f}\right) F(u)  = \frac{1-F'(u)}{F(u)} u_x^2 + u_{xx}
\end{eqnarray}
Now, we have a choice so that we can have approximation, which is if $\frac{f'}{f}$ is sufficiently close to 1 either $\forall t$ or as $t\uparrow$, we can neglect the term containing $(1- \frac{f'}{f})$ as it is comparatively very small. Following this approximation, we can eliminate the coefficient of $F(u)$. This approximation allows us to express $\frac{f'}{f} \approx 1$ so that we can represent, $f \approx e^t$. This reduces the equation \eqref{tru} to yield the following PDE, 
\begin{eqnarray}
\label{apprxu}
u_t  = \frac{1-F'(u)}{F(u)} u_x^2 + u_{xx}
\end{eqnarray}
This gives us a more closer bound to the original PDE than the one obtained by Ansatz type I i.e \eqref{trfeq}. The analytical proof of this can be done using Maximum principle and is given in the Appendix B. 

\subsubsection{Dispersion relation}

{\noindent}Let us consider linearized form of the function $F$ given by, $F(u) = u$, because this is the dominant contribution to the wave propagation as $0\le u\le 1$ . Hence, taking this dominant contribution in the front of the wave, the PDE \eqref{apprxu} reduces to the following simpler form,
\begin{eqnarray} 
\label{simpl}
u_t= u_{xx}
\end{eqnarray}
We followed the leading edge analysis technique proposed by Murray \cite{murray}. We assume an initial condition of the form given by,
\begin{eqnarray}
u(x,0) \sim e^{-ax} \quad \text{as} \quad x\rightarrow \infty 
\end{eqnarray}
where, $a$ is a constant. Then, following Murray's technique, we define the travelling wave ansatz given by, 
\begin{eqnarray}
\label{ans}
u(x,t) =u(x-ct)=A e^{-a(x - ct)}
\end{eqnarray} 
Now, putting the ansatz \eqref{ans} to the equation \eqref{simpl}, and after simplification, we get, $c = a$ which have the dimension of velocity.

\subsubsection{Application to Fisher's case}

{\noindent}Let us apply the proposed transformation technique to original Fisher's case, where, $F(u) = u(1-u)$. From this transformation equation \eqref{apprxu}, we get the following modified PDE and transformation function,
\begin{eqnarray} 
u_t &=& \frac{2}{1-u} u_x^2 + u_{xx} \\
G(u) &=& \frac{u}{1-u} = e^t h(x,t)
\end{eqnarray} 
A point to be noted about the diffusion equation $h_t = h_{xx} - h $ for $h(x,t)$ is that this can be analogous to the conduction of heat along a wire which loses heat from its surface at a rate proportional to its temperature \cite{crank} (p.329). This PDE has been extensively studied and solutions for various kinds are known \cite{carslaw,danck,sattinger,kamin,bricmont}. Now, for solving $h_t = h_{xx} - h $, we followed the procedure in \cite{carslaw} and we get the solution as $ h(x,t) = e^{-t} w(x,t) $, where, $w(x,t)$ now satisfies the diffusion equation $w_t = w_{xx}$. Thus, solving this diffusion equation, we obtain the solution $w(x,t)$ which can be related to $u(x,t)$ as given by,
\begin{eqnarray}
G(u) = \frac{u}{1-u} = w(x,t)\nonumber\\
u(x,t)= \frac {w(x,t)}{1+w(x,t)}
\label{uw}
\end{eqnarray}
To demonstrate this ansatz we chose an initial condition to be $u(x,0) = \frac{1}{1+e^{b x}}$ so that $w(x,0) = e^{-b x}$. From this initial condition, one can get $w(x,t)= e^{-b(x-bt)}$. Now, putting in the equation \eqref{uw}, we arrive at the solution $u$ as,
\begin{eqnarray}
u(x,t)= \frac {e^{-b(x-bt)}}{1+e^{-b(x-bt)}}
\end{eqnarray}

\subsubsection{Case B: h(x,t) satisfies $\displaystyle h_t= h_{xx} -\frac{2}{h}h_x^2$}

{\noindent}As a conjecture we study this particular case of transformation applicable only for the Fisher type of nonlinearities. We take the ansatz for $G$ in terms of $h(x,t)$ as given by, 
\begin{eqnarray}
G(u)= \frac{1}{1+h(x,t)}
\end{eqnarray} 
From this ansatz, we have,
\begin{eqnarray}
h_x&=& -(1+h)^2 G' u_x\nonumber\\
h_t&=& -(1+h)^2 G' u_t\nonumber\\
h_{xx} &=& \frac{2h_x^2}{1+h} -G'' (1+h)^2 u_x^2 -G' (1+h)^2 u_{xx}\nonumber
\end{eqnarray}
Then, substituting the above expressions into the equation $\displaystyle h_t= h_{xx} -\frac{2}{h}h_x^2$, we get the transformed PDE as,
\begin{eqnarray}
u_t  = \frac{G''}{G'} u_x^2 + u_{xx}
\end{eqnarray} 
Now, as an adhoc approximation we can either neglect the term containing $\frac{G''}{G'} u_x^2 $ which is clearly not acceptable, and hence, we have to proceed as follows. Now if we take $G(u) = u$ we can eliminate the term $\frac{G''}{G'}u_x^2 $ completely and thus obtain a completely linearised homogeneous heat equation. The problem is thus reduced to solving the PDE $\displaystyle h_t= h_{xx} -\frac{2}{h}h_x^2$ which to the best of our knowledge appears to be not known and the only mention of this PDE is in the work of \cite{puri2}(Eq.6 in p.455).

\section{Conclusion}

{\noindent}The spatiotemporal species population growth and development is quite complicated process driven by a large number of factors. Certain aspects of such complicated processes and dynamics can be well explained in a simplified manner by the classic Fisher-KPP PDEs. However, the functional form $F$ in population genetics can have various forms which are generally nonlinear in nature to deal with different aspects of species evolution and dynamics. We have extensively reviewed various methodological perspectives of solving Fisher-KPP types of PDEs and both analytical and numerical results along this direction.\\

{\noindent}In this work, we proposed a systematic nonlinear transformation method which beautifully converts the original classic Fisher-KPP type class of PDEs into exactly solvable class of PDEs. Further, the significance of this method is that this method inherits essential inherent properties, such as, relaxation, asymptotic properties etc from the original non-linear dynamical system which we proved in this work. Further, we also showed that this particular transformation method allows us to obtain exact closed form solution of the nonlinear Fisher-KPP type PDEs in terms of the heat kernel and analytical approximations through the two variable Hermite polynomials. With our method we verified some of the numerical results of these PDEs studied before.\\ 

{\noindent}We considered four specific functional forms of $F$ with special connection with the population genetics in the original Fisher-KPP type PDEs and applied our method to address some of the important properties and dynamics of evolution of the mutant gene. We then calculate the front velocities of the propagating wave and found a power law relaxation for the front velocities. On the basis of the above work we put together two hypothesis: Firstly, together with more advanced renormalization group theories we can get better bounds and secondly on modifying the growth functions i.e. the exponential we hope to predict some new interesting phenomena. We hope these ideas and proposed method will open up new areas of further rigorous research into these techniques.\\

\noindent\textbf{Acknowledgement:} PM is financially supported by Council of Scientific and Industrial Research (CSIR), India. RKBS acknowledges financial support from Department of Science and Technology (DST), India under Matric scheme.


\section{Appendices}
\subsection{Heat kernel and its characteristics}

\subsubsection{Estimates of the heat kernel}

\noindent Our nonlinear transform extensibly uses the heat equation throughout because of the known properties of the heat kernel. For a ready reference of a particular property which makes the homogenous heat equation a particular suitable candidate  we cite a lemma from Moet \cite{moet}.\\

\noindent \textbf{Lemma 1.} \textit{Let $h$ be the solution of the initial valued problem, 
\begin{eqnarray}
h_t = h_{xx}, \quad x \in \mathbb{R} \text{ and } t > 0
\end{eqnarray}
with the initial condition, $h(x, 0) = h_0(x), \quad x \in \mathbb{R}$, where, $h_0$ is a bounded continuous function having at most a finite number of discontinuity points, which belongs to \( L^p(\mathbb{R}) \) for some \( p \geq 1 \). Then $h$ satisfies the estimates }
\begin{eqnarray}
|h(x, t)| &\leq &(4\pi t)^{-1/2} \|h_0\|_{L^p(\mathbb{R})} \quad \text{for all } (x, t) \in \mathbb{R} \times (0, \infty)\\
\|h(\cdot, t)\|_{L^q(\mathbb{R})} &\leq &(4\pi t)^{-(1/2)(1/p - 1/q)} \|h_0\|_{L^p(\mathbb{R})} \quad \text{for all } t > 0 \text{ and any } q \geq p \geq 1
\end{eqnarray}

\noindent\textbf{Proof:} \textit{The proof can be found in \cite{moet}.}\\

{\noindent}This is simply stating a known fact about how the heat provided initially is transported everywhere in the domain. This is a standard estimate using the parabolic inequalities and for the proof we refer to \cite{moet} .This result is especially useful because this states the exact way the solution $h(x,t)$ depends on its initial values. The above lemma also demonstrates the regularisation features of the heat kernel that makes the solutions converge uniformly to their asymptotic value and we will be needing this behaviour crucially for the nonlinear transform to work.

\subsubsection{Solution using Heat Kernel in the nonlinear transformation}

\noindent \textbf{\textit{Proposition 2:}} \textit{The general closed form solution is given by the transformation G as:}
\begin{eqnarray}
\label{Heqnsol}
G[u(x,t)]=\left[\frac{f(t)}{f(0)\sqrt{4 \pi t}} \int_{\mathbb{R}} G(u(x',0)) e^{- \frac{(x-x')^2}{4t} } dx'\right] 
\end{eqnarray}

\noindent\textbf{\textit{Proof: }}
We start by stating the fundamental solution of the heat equation which is given by,
\begin{eqnarray}
\label{hkr}
h(x,t) = \frac{1}{\sqrt{4 \pi t}} \int_{\mathbb{R}}h(x',0) e^{- \frac{(x-x')^2}{4t} } dx'
\end{eqnarray}
The initial condition of the heat hernel can be obtained from the transformation equation \eqref{trans} by putting $t=0$ which is given by,
\begin{eqnarray}
\label{hki}
h(x,0)=\frac{1}{f(0)}G[u(x,0)]
\end{eqnarray}
with $f(0) \neq 0$. Combining these two equations \eqref{hkr} and \eqref{hki}, we have,
\begin{eqnarray}
G[u(x,t)]=\left[\frac{f(t)}{f(0)\sqrt{4 \pi t}} \int_{\mathbb{R}} G[u(x',0)] e^{- \frac{(x-x')^2}{4t} } dx'\right]
\end{eqnarray}
Which when used with the inverse transform of $G^{-1}$ gives us the solution $u(x,t)$.

\subsubsection{Solutions with multivariable Hermite polynomials }

\noindent\textbf{\textit{Proposition 3:}} \textit{The complete series solution for u(x,t) can be obtained with arbritrary precision by,}
\begin{eqnarray}
 G(u(x,t)) = \left\{ \frac{f(t)}{f(0)} \sum_{n=0}^{\infty} \frac{\left[G(u(x,0))\right]^{(n)}}{n!} {}^{(2)}H_n(x,t) \right\}
\end{eqnarray}

\noindent\textbf{\textit{Proof:}} We start by stating that the multi-variable Hermite polynomials satisfy the following PDE, 
\begin{equation}
\frac{\partial}{\partial t}{}^{(2)}H_n(x,t)= \frac{\partial^{2}}{\partial x^{2}} {}^{(2)}H_n(x,t)
\end{equation} 
where the multi-variable Hermite polynomials is given in terms of ordinary Hermite polynomials $H_n$ as given by,  
\begin{eqnarray}
{}^{(2)}H_n(x,t) = n!\sum_{r=0}^{[n/2]} \frac{H_{(n-2r)}(x) H_r(t)}{n(n-2r)!(r)!}
\end{eqnarray}
Then, the heat kernel can be expressed in terms of Hermite polynomials given by, 
\begin{eqnarray}
h(x,t) = \sum_{n=0}^{\infty} \frac{h^{(n)}(x,0)}{n!} {}^{(2)}H_n(x,t) 
\end{eqnarray}
where, $h^{(n)}(x,0)$ denotes the nth derivative. Thus again by using the transformation function $G$ in equation \eqref{hki}, we get,
\begin{eqnarray}
G[u(x,t)] &=& f(t) \sum_{n=0}^{\infty} \frac{h^{(n)}(x,0)}{n!} {}^{(2)}H_n(x,t)\nonumber\\
&=& \left\{ \frac{f(t)}{f(0)} \sum_{n=0}^{\infty} \frac{\left[G(u(x,0))\right]^{(n)}}{n!} {}^{(2)}H_n(x,t) \right\}
\end{eqnarray}
where, $[\quad ]^{(n)}$ denotes the nth derivative with respect to $x$.

\subsection{Proofs of Comparsion principle }

\noindent \textbf{\textit{Proof:}} We begin by defining $v(x,t)$ which satisfies PDE \eqref{trfeq} and $u$ in equation \eqref{apprxu} which is given by,
\begin{eqnarray}
u_t  = \frac{1-F'(u)}{F(u)} u_x^2 + u_{xx}
\end{eqnarray}
We state the following parabolic inequalities together with a set of conditions : 
\begin{enumerate}
    \item $v(x,t), u(x,t) \in [0,1] \quad \forall x,t \in \mathbb{R} \times \mathbb{R}^+$ where $u(x,t)$ satisfies above PDE and $v(x,t)$ which satisfies PDE \eqref{trfeq}
    \item $v_t - v_{xx} - H(v,v_x)  \geq u_t - u_{xx} - H(u, u_x) $  where $H(u,u_x) = \left( \frac{1 - F'(u(x,t)) }{F(u(x,t))} \right) (u_x(x,t))^2$ and  $H(v,v_x) =\left( \frac{1 - F'(v(x,t)) }{F(v(x,t))} \right) (v_x(x,t))^2$ remain uniformly bounded in $\mathbb{R} \times [0,T]$ with $T<\infty $.
    \item $1 \geq v(x,0) \geq u(x,0) \geq 0.$
\end{enumerate}
These three conditions then imply
\begin{equation}
\label{ineq}
    v(x,t) \geq u(x,t) \quad \forall \; (x,t) \in \mathbb{R} \times [0,T], T<\infty 
\end{equation} 
which is our desired proposition.

{\noindent}To prove the above inequality \eqref{ineq} we first need some estimates. We get these estimates by using  \textbf{Mean Value Theorem}:
\begin{equation}
\label{mvt1}
   \frac{ \left( \frac{1 - {F'(v)}}{F(v)} \right) - \left( \frac{1 - {F'(u)}}{F(u)} \right)} {(v-u)}  \\
= \left[ \frac{ [F'(w)]^2 - F'(w) -F''(w)F(w)}{[F(w)]^2} \right] 
\end{equation}
where, $w$ is numerically between $u$ and $v$. This equation \eqref{mvt1} further helps us to estimate,
\begin{eqnarray}
H(v, v_x) - H(u, u_x) &=& \frac{v_x^2 \left( 1 - F'(v) \right)}{F(v)} - \frac{u_x^2 \left( 1 - F'(u) \right)}{F(u)} \nonumber\\ 
&=& \frac{v_x^2 \left( 1 - F'(v) \right)}{F(v)} - \frac{u_x^2 \left( 1 - F'(v) \right)}{F(v)} -\frac{u_x^2 \left( 1 - F'(u) \right)}{F(u)}  +\frac{u_x^2 \left( 1 - F'(v) \right)}{F(v)}   \nonumber\\
&=& \frac{(v_x^2-u_x^2) \left( 1 - F'(v) \right)}{F(v)} + u_x^2 \left( \frac{1 - {F'(v)}}{F(v)} - \frac{1 - {F'(u)}}{F(u)} \right)\nonumber\\
&=& \frac{(v_x^2-u_x^2) \left( 1 - F'(v) \right)}{F(v)} \nonumber\\
&&+ u_x^2 \left[ \frac{ [F'(w)]^2 - F'(w) -F''(w)F(w)}{[F(w)]^2} \right] (v - u)
\end{eqnarray}
Now, for the condition $w > u$, we obtain the estimate given by the following inequality,
\begin{equation}
\label{ineq1}
H(v, v_x) - H(u, u_x) \geq  \frac{(v_x^2-u_x^2) \left( 1 - F'(v) \right)}{F(v)} + u_x^2 \left[ \frac{ [F'(u)]^2 - F'(u) -F''(u)F(u)}{[F(u)]^2} \right] (v - u) 
\end{equation}
Further, we define a function $\Phi$ by, 
\begin{eqnarray}
\label{phiu}
\Phi(u) = \left[ \frac{ [F'(u)]^2 - F'(u) -F''(u)F(u)}{[F(u)]^2} \right]
\end{eqnarray} 
Then, the inequality \eqref{ineq1} becomes, 
\begin{eqnarray}
\label{hhinq}
H(v, v_x) - H(u, u_x) \geq  \frac{(v_x^2-u_x^2) \left( 1 - F'(v) \right)}{F(v)} + (v - u) \Phi(u) u_x^2 
\end{eqnarray}
Now, we use these growth estimates to check how the difference between $v$ and $u$ changes with time. We assert that as $ t \to \infty $ this difference $ (v - u) \rightarrow 0 $  as both $v$ and $u$ tends to the steady state. To prove this statement, we define a new quantity $z=(v-u)e^{-Ct}$, where, the magnitude of $C$ is determined by the comparison of the new term in the transformed PDE. We also have asserted that the quantity $ \frac{ v_x (1 - F'(v) ) }{F(v)}  $ remains uniformly bounded in $\mathbb{R} \times [0,T]$ with $T<\infty $. Hence, there exists \( M_T < \infty\), such that,
\begin{equation}
\sup_{(x,t) \in \mathbb{R} \times [0,T]} \left[\frac{ v_x (1 - F'(v) ) }{F(v)} ,  \frac{ u_x (1 - F'(u) ) }{F(u)}  \right]  = M_T
\end{equation}
This yields $C= 1+2M_T^2$ which gives $z=(v-u)e^{-(1+2M_T^2)t}$. Further, we also know from the condition 2 asserted above that is given by,
 \begin{eqnarray}
 (v - u)_t - (v - u)_{xx} \geq H(v, v_x) - H(u, u_x)
 \end{eqnarray}
Now, let us consider the following PDE in $z=(v-u)e^{-(1+2M_T^2)t}$, 
\begin{eqnarray}
\Bigg[z_t - z_{xx} - z_x \left( \frac{(v_x+u_x) \left( 1 - F'(v) \right)}{F(v)}  \right) \Bigg]e^{(1+2M_T^2)t}
&=& (v - u)_t - (1 + 2M_T^2)(v - u)- (v - u)_{xx} \nonumber\\
&&- (v - u)_x \left[ \frac{(v_x+u_x) \left( 1 - F'(v) \right)}{F(v)}  \right]
\end{eqnarray}
Further, using the condition 2 stated above and then incorporating equation \eqref{hhinq}, we have,
\begin{eqnarray}
\Bigg[z_t - z_{xx} - z_x \left( \phi_1\right) \Bigg]e^{(1+2M_T^2)t}
&\geq& H(v, v_x) - H(u, u_x) \nonumber\\
&&-(v - u)_x \left( \frac{(v_x+u_x) \left( 1 - F'(v) \right)}{F(v)}  \right)- (1 + 2M_T^2)(v - u) \nonumber\\
&\geq& \frac{(v_x^2-u_x^2) \left( 1 - F'(v) \right)}{F(v)} 
 - \frac{(v_x^2-u_x^2) \left( 1 - F'(v) \right)}{F(v)} \nonumber\\
&&+ u_x^2 \left[ \Phi(u) \right] (v - u) - (1 + 2M_T^2)(v - u) \nonumber\\
&\geq&  u_x^2 \left[ \Phi(u) \right] (v - u) - (1 + 2M_T^2)(v - u)   \nonumber\\
&\geq& \left[-(1 + 2M_T^2)+ u_x^2 \left[ \Phi(u) \right]  \right] (v - u) \nonumber\\
&\geq&  \left[-(1 + 2M_T^2)+ u_x^2 \left[ \Phi(u) \right]  \right] ze^{(1 + 2M_T^2)t}\nonumber\\
&\geq&\phi_2 ze^{(1 + 2M_T^2)t}
\end{eqnarray}
so that we can simplify the above expression as the following inequality,
\begin{eqnarray}
z_t - z_{xx} - \left[\phi_1\right]z_x-\left[\phi_2\right]z \geq 0
\end{eqnarray}
where $\phi_1 = \left( \frac{(v_x+u_x) \left( 1 - F'(v) \right)}{F(v)}  \right) $ and $ \phi_2= \left[-(1 + 2M_T^2)+ u_x^2 \left[ \Phi(u) \right]  \right]$ with $\Phi(u)$ as declared above in equation \eqref{phiu}. As both $\phi_1$ and $ \phi_2$ are bounded $\forall (x,t) \in \mathbb{R} \times [0,T]$ this proves by Maximum Principle for parabolic systems (Lemma 2, Page 166, Chap. 3 of \cite{hans}) that $z \geq 0 \quad \forall (x,t) \in \mathbb{R} \times [0,T]$ which completes the proof of our proposition.

\subsection{Calculation of asymptotic values of wave velocities of Fisher-KPP equation}

\subsubsection{Fisher Case}

{\noindent}We have the wave velocity for $w=\frac{3}{4}$,
\begin{eqnarray}
V_{w=\frac{3}{4}}=  \frac{2 \log \left(\frac{\eta a}{3(1- \eta ) \sqrt{a^2+2 t}}\right)+a^2+4 t-1}{ \sqrt{2\left(a^2+2 t\right) \left(\log \left(\frac{\eta a}{3(1-\eta ) \sqrt{a^2+2 t}}\right)+t\right)}}
\end{eqnarray}
On calculating for the limit $t\rightarrow\infty$, we get, $\displaystyle\lim_{{t \to \infty}} V_{w=\frac{3}{4}} =2$.\\

{\noindent}Similarly for $w=\frac{1}{4}$, we have, 
\begin{eqnarray}
V_{w=\frac{1}{4}}=   \frac{2 \log \left(\frac{3\eta a}{(1- \eta ) \sqrt{a^2+2 t}}\right)+a^2+4 t-1}{ \sqrt{2\left(a^2+2 t\right) \left(\log \left(\frac{3\eta a}{(1- \eta ) \sqrt{a^2+2 t}}\right)+t\right)}}
\end{eqnarray}
and thus we verify the asymptotic value of the velocity is $\displaystyle\lim_{{t \to \infty}} V_{w=\frac{1}{4}} =2$.

\subsubsection{Heterozygote Inferior case}

{\noindent}We solve to get for $u(x,t) = w= 0.99$ the front position corresponding to this value of w as 
\begin{eqnarray}
x= \sqrt{2(a^2 + 2t)\left( t +  \ln\left( \frac{3\eta a}{4(1-\eta) \sqrt{a^2 + 2t}} \right)\right)}
\end{eqnarray}
Thus we calculated the velocity as,
\begin{eqnarray}
V_{w=0.99}=  \frac{2 \log \left(\frac{ 99 \eta a}{9604(1- \eta ) \sqrt{a^2+2 t}}\right)+a^2+4 t-1}{ \sqrt{2\left(a^2+2 t\right) \left(\log \left(\frac{  99\eta a}{9604(1-\eta ) \sqrt{a^2+2 t}}\right)+t\right)}}
\end{eqnarray}
We verify again the asymptotic value of the wave velocity as we have obtained above $\displaystyle\lim_{{t \to \infty}}V_{w=0.99}=2 $



\section{FIGURES}

\end{document}